\documentclass[english,twocolumn]{revtex4-1}
\usepackage[T1]{fontenc}
\usepackage[latin9]{inputenc}
\usepackage{color}
\usepackage{babel}
\usepackage{float}
\usepackage{textcomp}
\usepackage{amsbsy}
\usepackage{amstext}
\usepackage{amssymb}
\usepackage{amsmath}
\usepackage{graphicx}
\usepackage{esint}
\PassOptionsToPackage{normalem}{ulem}
\usepackage{ulem}
\usepackage[unicode=true]
 {hyperref}

\makeatletter

\providecommand{\tabularnewline}{\\}

\DeclareMathOperator{\tr}{Tr}
\newcommand{\Id}[1][]{\mathbb{I}_{#1}}
\usepackage[braket, qm]{qcircuit}
\usepackage[subrefformat=parens,labelformat=parens]{subfig}
\usepackage{caption}

\newcommand{\prepareC}[1]{*{\xy*+=+<.5em>{\vphantom{#1\rule{0em}{.1em}}}*\cir{l^r};p\save*!L{#1} \restore\save+UC;+UC+<.5em,0em>*!L{\hphantom{#1}}+R **\dir{-} \restore\save+DC;+DC+<.5em,0em>*!L{\hphantom{#1}}+R **\dir{-} \restore\POS+UC+<.5em,0em>*!L{\hphantom{#1}}+R;+DC+<.5em,0em>*!L{\hphantom{#1}}+R **\dir{-} \endxy}}

\newcommand{\ignore}[1]{}

\makeatother

\begin{document}
\title{Visualizing Kraus operators for dephasing noise during application
of the $\sqrt{\mathrm{\mathrm{SWAP}}}$ quantum gate}
\author{Nicolas Andr\'{e} da Costa Morazotti}
\affiliation{S\~{a}o Carlos Institute of Physics, University of S\~{a}o Paulo, PO Box 369,
13560-970, S\~{a}o Carlos, SP, Brazil}
\author{Reginaldo de Jesus Napolitano}
\affiliation{S\~{a}o Carlos Institute of Physics, University of S\~{a}o Paulo, PO Box 369,
13560-970, S\~{a}o Carlos, SP, Brazil}
\begin{abstract}
We consider the case of a $\sqrt{\mathrm{SWAP}}$ quantum gate and
its optimized entangling action, via continuous dynamical decoupling,
in the presence of dephasing noise. We illustrate the procedure in
the specific case where only the two-qubit operation is controlled
and no single-qubit operations are included in the description. To
compare the optimized dynamics in the presence of noise with the ideal
case, we use the standard fidelity measure. Then we discuss the importance
of using optimized gates in the quantum operational-probabilistic
theory. Because of their importance for the explicit construction
of the completely positive maps representing the operations, we derive 
optimized Kraus operators in this specific case, focusing on the
entanglement operation. We then show how to visualize the time
evolution of each Kraus operator as a curve in a three-dimensional
Euclidean space. Finally, we connect this formalism with the operational 
framework of quantum mechanics by describing a possible set of measurements
that could be performed to obtain the Kraus operators.
\end{abstract}
\maketitle

\section{Introduction}

Optimal control has been a regularly studied subject for a long time,
mainly for studies on spin dynamics~\cite{Khaneja2005},
and it has become increasingly more relevant in the context of the
quantum information science~\cite{PhysRevA.85.052327,Green_2013,PhysRevA.102.052203}.
A very exciting prospect has been already envisaged starting in the
first decade of the present century: the development of a quantum
internet~\cite{Kimble2008,Ritter2012,quantum-net}. Recent experimental
efforts have managed to establish a quantum-communication network
over 4600 km~\cite{Chen2021}. In this context of quantum networking
and communications, important topics have been referred to as transduction~\cite{Mirhosseini2020}
and transfer of entanglement~\cite{Budich_2010,PhysRevA.98.062329}.
In particular, flying qubits which get entangled on the fly with target
qubits with which they interact have been considered in the context
of scattering~\cite{Banchi2017quantumgatesbetween}. Inspired by these
basic investigations, realistic implementation of distribution of
entanglement over quantum networks is already occurring~\cite{PhysRevLett.125.260506,PhysRevLett.126.020503}.
However, all these possibilities are plagued by the unavoidable noise
causing decoherence, which limits immensely the progress of implementation
of quantum technologies, mainly when the carriers of quantum information
are not photons~\cite{Bogdanov532}. It is, therefore, indispensable
to understand the effects of quantum noise in quantum-information
processing, including ways to correct for errors or to prevent them
from happening~\cite{RevModPhys.87.307}.

Any quantum gate able to entangle two qubits together with general
single-qubit operations will suffice for universal quantum processing~\cite{michaelnielsen2011}.
Realization of entanglement between pairs of qubits by approaching
and subsequently splitting them has been recently demonstrated using
lattice surgery~\cite{Erhard2021}. Also, addressability of individual
qubits has even been extended to chromium molecules~\cite{Bayliss1309}.
Indistinguishability symmetry has been explored in the context of
entanglement and quantum control~\cite{Nosrati2020,PhysRevLett.125.180402,PhysRevLett.125.230502}.
Along the lines of these particular recent developments, the $\sqrt{\mathrm{SWAP}}$
gate has long been realized in the laboratory using rubidium atoms
in optical lattices, promising control over the effective exchange
interaction~\cite{Anderlini2007}. Based on this possibility of articulating
the time dependence of the effective interaction between two qubits,
here we approach the case of quantum control of a $\sqrt{\mathrm{SWAP}}$
entangling quantum gate in the presence of dephasing noise, which
we simulate by introducing a boson bath. Relaxation times due to amplitude
damping, because it involves transfer of energy by dissipation, are
usually much longer than pure-decoherence times~\cite{quiroga}. We,
therefore, focus on the situation where we have the complete action
of the $\sqrt{\mathrm{SWAP}}$ gate fast compared to relaxation, but
slow compared to dephasing.

We optimize the time dependence during the application of the entangling
gate, while noise is present, by the well-studied continuous
dynamical-decoupling procedure~\cite{PhysRevA.75.022329}. We compare the
optimized noisy dynamics with the ideal, noiseless one by calculating
the fidelity measure of our output operation. With the optimization of
the gate under the perturbation due to coupling with the environment, we
derive a set of corresponding Kraus
operators~\cite{kraus1983,KEYL2002431}. These operators provide the
open-system evolution of any initial reduced matrix and are required to
establish the completely positive maps representing the quantum
operations in an Operational-Probabilistic Theory (OPT) of quantum
mechanics~\cite{PhysRevA.81.062348,PhysRevA.84.012311,Oreshkov2015}.
Given the importance of the Kraus operators in the open-system dynamics
of quantum-information processing, we present a prescription to
visualize them via a one-to-one correspondence between each Kraus
operator and a three-dimensional real vector. The whole dynamics as
embedded in the time dependence of each Kraus operator is, thus,
transcribed into the trajectory described by each of the corresponding
vectors. Although our focus is on the $\sqrt{\mathrm{SWAP}}$ gate under
dephasing, we believe the visualization prescription may serve to
inspire its generalization to more complex situations, analogously to
the recent extension of the Bloch-sphere and Bloch-vector concepts to
the case of a qutrit~\cite{PhysRevA.93.062126} and, more generally, of a
qudit~\cite{Bertlmann_2008}.

In Sec. \ref{sec:Optimization-of-the} we describe the optimization of
the $\sqrt{\mathrm{SWAP}}$ quantum gate in the presence of dephasing
noise. In Sec. \ref{sec:numer-simul} we present our numerical
calculations for the optimization of the time-dependence of the
controlled entanglement, showing the fidelity measure as a function of
several parameters involved in the simulations. In
Sec. \ref{sec:visu-kraus-oper} we calculate the optimized Kraus
operators and present a prescription to their three-dimensional
visualization. In Sec. \ref{sec:opts-quantum-process}, in the spirit of 
Quantum Process Tomography (QPT)~\cite{qpt}, we present a possible set of measurements 
illustrating how to obtain the process matrix~\cite{michaelnielsen2011} in the OPT language, which we briefly review in Appendix \ref{sec:opts}.
In Sec. \ref{sec:conclusion} we
conclude our analysis and its results.

\section{\label{sec:Optimization-of-the}Optimization of the $\sqrt{\mathrm{SWAP}}$
quantum gate}

The SWAP quantum gate has been considered both theoretically and experimentally~\cite{Anderlini2007}
and, together with universal one-qubit operations, can be used to
implement universal quantum computing~\cite{michaelnielsen2011}.
Giving its importance, we investigate its optimal application by means
of controlling its strength as a function of time to be optimized,
as the experimental implementation of Ref.~\cite{Anderlini2007} has
given evidence to its feasibility. Thus, we consider the gate Hamiltonian
given by
\begin{eqnarray}
H_{G}\left(t\right) & = & \hbar\Omega\left(t\right)G,\label{HG}
\end{eqnarray}
where $G$ is an operator that acts on the two qubits, as we
specify below, and $\Omega\left(t\right)$ is assumed to be a controlled
frequency whose time dependence we can choose arbitrarily.
Let us consider dephasing noise by introducing a boson field~\cite{quiroga}.
In this case, the total spin-boson Hamiltonian, much studied in the
dissipative case~\cite{PhysRevLett.46.211,RevModPhys.59.1}, can be
written as
\begin{eqnarray}
H\left(t\right) & = & H_{G}\left(t\right)+\hbar\sum_{s}\omega_{s}b_{s}^{\dagger}b_{s}\nonumber \\
 &  & +\hbar\sum_{k=1}^{2}\sigma_{k,z}\sum_{s}\left(g_{k,s}b_{s}+g_{k,s}^{\ast}b_{s}^{\dagger}\right),\label{H}
\end{eqnarray}
where $\omega_{s}$ is the frequency of the boson in its $s$th mode,
$b_{s}$ and $b_{s}^{\dagger}$ are the annihilation and creation
operators of a boson quantum in the $s$th mode, $\sigma_{k,z}$ is
the $z\text{-axis}$ Pauli matrix of the $k$th qubit, and $g_{k,s}$
is the coupling constant of the $k$th qubit with the $s$th boson
degree of freedom. For concreteness, as we have emphasized above,
let $G$ describe the SWAP gate:
\begin{eqnarray}
G & = & \left|0\right\rangle \left\langle 0\right|\otimes\left|0\right\rangle \left\langle 0\right|+\left|1\right\rangle \left\langle 1\right|\otimes\left|1\right\rangle \left\langle 1\right|\nonumber \\
 &  & +\left|0\right\rangle \left\langle 1\right|\otimes\left|1\right\rangle \left\langle 0\right|+\left|1\right\rangle \left\langle 0\right|\otimes\left|0\right\rangle \left\langle 1\right|\nonumber \\
 & = & \frac{1}{2}\mathbb{I}_{S}+\frac{1}{2}\sigma_{1,x}\sigma_{2,x}+\frac{1}{2}\sigma_{1,y}\sigma_{2,y}+\frac{1}{2}\sigma_{1,z}\sigma_{2,z}\nonumber \\
 & = & \frac{\mathbb{I}_{S}+\boldsymbol{\sigma}_{1}\boldsymbol{\cdot}\boldsymbol{\sigma}_{2}}{2}.\label{G}
\end{eqnarray}
Hence, it is easy to verify that
\begin{eqnarray}
G^{2} & = & \mathbb{I}_{S},\label{Gsquared}
\end{eqnarray}
where $\mathbb{I}_{S}$ is the two-qubit identity operator. Here,
the states of each qubit are given by the eigenstates of the Pauli operator
for the $z$ axis, forming the logical basis set: $\left\{ \left|0\right\rangle ,\left|1\right\rangle \right\} .$
Conventionally, in the third line of Eq. (\ref{G}) we adopt the indices
$k=1,2$ as referring to the qubits whose operator factors in the
tensor products are on the left and on the right, respectively.

In the absence of noise, that is, choosing $g_{k,s}=0$ in Eq. (\ref{H}),
the simplest choice of $\Omega\left(t\right)$ that takes the factorized
state $\left|0\right\rangle \otimes\left|1\right\rangle ,$ after
a time interval $\tau,$ to a target state that is maximally entangled,
namely,
\begin{eqnarray}
\left|\psi\left(\tau\right)\right\rangle  & = & \frac{1}{\sqrt{2}}\left|0\right\rangle \otimes\left|1\right\rangle -\frac{i}{\sqrt{2}}\left|1\right\rangle \otimes\left|0\right\rangle ,\label{ideal0}
\end{eqnarray}
 is simply a constant: $\Omega\left(t\right)=\pi/4\tau.$

Instead of the notation of Eq. (\ref{G}) for the two-qubit kets and
bras, henceforth we use a more compact notation for two-qubit logical
kets in terms of their counterpart physical kets, namely, $\left|2m+n\right\rangle \equiv\left|m\right\rangle \otimes\left|n\right\rangle $
for $m,n\in\left\{ 0,1\right\} .$ The interaction between the qubits
and the bosons in Eq. (\ref{H}) does not couple the subspace spanned
by $\left\{ \left|0\right\rangle ,\left|3\right\rangle \right\} $
to the one spanned by $\left\{ \left|1\right\rangle ,\left|2\right\rangle \right\} .$
Furthermore, $G$ only entangles the input states $\left|1\right\rangle $
and $\left|2\right\rangle $ in this new logical notation, that is,
if we start with initial logical state $\left|0\right\rangle $
or $\left|3\right\rangle ,$ the dynamics driven by $G,$ besides
not leaving the subspace spanned by $\left\{ \left|0\right\rangle ,\left|3\right\rangle \right\} ,$
will not generate a superposition between $\left|0\right\rangle $
and $\left|3\right\rangle .$ Therefore, since our purpose is to study
optimized entangling evolutions, we just need to consider the logical
subspace spanned by $\left\{ \left|1\right\rangle ,\left|2\right\rangle \right\} $
and, for concreteness, let us define the starting two-qubit state,
at $t=0,$ as the one whose logical representation is given by the
ket $\left|1\right\rangle .$ We want to have an ideal superposition
after gate $G$ acts for a total time interval $\tau,$ which is the
same as the one of Eq. (\ref{ideal0}), but that in the compact notation,
in terms of the logical qubit, now reads:
\begin{eqnarray}
\left|\psi\left(\tau\right)\right\rangle  & = & \frac{1}{\sqrt{2}}\left|1\right\rangle -\frac{i}{\sqrt{2}}\left|2\right\rangle .\label{ideal}
\end{eqnarray}
Restricted to the relevant subspace spanned by $\left\{ \left|1\right\rangle ,\left|2\right\rangle \right\} ,$
we have reduced the two-qubit physical system to a logical single-qubit
system coupled with dephasing noise, so that now we can use, for $G,$
the Pauli matrix $\sigma_{x}$ acting on the logical basis as the
exchange operator:
\begin{eqnarray}
\sigma_{x}\left|1\right\rangle  & = & \left|2\right\rangle \label{sigmax1}
\end{eqnarray}
and
\begin{eqnarray}
\sigma_{x}\left|2\right\rangle  & = & \left|1\right\rangle .\label{sigmax2}
\end{eqnarray}
In this simpler formulation, Eq. (\ref{H}) is now mapped to the following
total spin-boson Hamiltonian:
\begin{eqnarray}
H_{S}\left(t\right) & = & \hbar\Omega\left(t\right)\sigma_{x}+\hbar\sum_{s}\omega_{s}b_{s}^{\dagger}b_{s}\nonumber \\
 &  & +\hbar\sigma_{z}\sum_{s}\left(g_{s}b_{s}+g_{s}^{\ast}b_{s}^{\dagger}\right),\label{HS}
\end{eqnarray}
where
\begin{eqnarray}
\sigma_{z}\left|1\right\rangle  & = & \left|1\right\rangle ,\label{sigmaz1}
\end{eqnarray}
\begin{eqnarray}
\sigma_{z}\left|2\right\rangle  & = & -\left|2\right\rangle ,\label{sigmaz2}
\end{eqnarray}
and we have defined
\begin{eqnarray}
g_{s} & \equiv & g_{1,s}-g_{2,s}.\label{gs}
\end{eqnarray}

To treat the effects of dephasing on the gate action, we now change
to the interaction picture, that is, we transform from the global
Schr\"{o}dinger density-matrix operator $\rho\left(t\right)$ to its interaction-picture
counterpart:
\begin{eqnarray}
\rho_{I}\left(t\right) & \equiv & U_{0}^{\dagger}\left(t\right)\rho\left(t\right)U_{0}\left(t\right),\label{PsiI}
\end{eqnarray}
where
\begin{eqnarray}
U_{0}\left(t\right) & \equiv & \exp\left[-i\Phi\left(t\right)\sigma_{x}\right]\exp\left(-i\sum_{s}\omega_{s}tb_{s}^{\dagger}b_{s}\right),\label{U0}
\end{eqnarray}
with
\begin{eqnarray}
\Phi\left(t\right) & \equiv & \int_{0}^{t}dt^{\prime}\,\Omega\left(t^{\prime}\right).\label{Phi}
\end{eqnarray}
It is now important to notice that $\rho\left(t\right)$ and $\rho_{I}\left(t\right)$
are meant to represent density-matrix operators of the whole system,
that is, the system that is composed by the logical qubit and the
thermal boson bath. Although we consider the initial qubit state as
pure, given by
\begin{eqnarray}
\rho_{S}\left(0\right) & = & \left|\psi\left(0\right)\right\rangle \left\langle \psi\left(0\right)\right|,\label{rhoS(0)}
\end{eqnarray}
usually we take the initial state of the thermal bath, since it is thermal, as a mixed
state, whose best description is in terms of its canonical density-matrix
operator, namely,
\begin{eqnarray}
\rho_{B}\left(0\right) & = & \frac{\exp\left(-\beta\hbar\sum_{s}\omega_{s}b_{s}^{\dagger}b_{s}\right)}{Z},\label{thermal}
\end{eqnarray}
where
\begin{eqnarray}
\beta & \equiv & \frac{1}{k_{B}T},\label{beta}
\end{eqnarray}
$T$ is the boson-bath temperature, $k_{B}$ is Boltzmann constant, and
\begin{eqnarray}
Z & \equiv & \mathrm{Tr}_{B}\left[\exp\left(-\beta\hbar\sum_{s}\omega_{s}b_{s}^{\dagger}b_{s}\right)\right]\label{partition}
\end{eqnarray}
is the partition function.

In the interaction picture, the corresponding evolution of the global
density-matrix operator is given by
\begin{eqnarray}
i\hbar\frac{d}{dt}\rho_{I}\left(t\right) & = & \left[H_{I}\left(t\right),\rho_{I}\left(t\right)\right],\label{evol}
\end{eqnarray}
with
\begin{eqnarray}
H_{I}\left(t\right) & \equiv & U_{0}^{\dagger}\left(t\right)H_{int}U_{0}\left(t\right)\nonumber \\
 & = & \hbar\sigma_{z}\left(t\right)\sum_{s}\left[g_{s}b_{s}\exp\left(-i\omega_{s}t\right)\right.\nonumber \\
 &  & +\left.g_{s}^{\ast}b_{s}^{\dagger}\exp\left(i\omega_{s}t\right)\right],\label{HI}
\end{eqnarray}
where

\begin{eqnarray}
\sigma_{z}\left(t\right) & \equiv & \left[i\Phi\left(t\right)\sigma_{x}\right]\sigma_{z}\exp\left[-i\Phi\left(t\right)\sigma_{x}\right]\nonumber \\
 & = & \sigma_{z}\cos\left[2\Phi\left(t\right)\right]+\sigma_{y}\sin\left[2\Phi\left(t\right)\right].\label{sigma(t)}
\end{eqnarray}
The interaction-picture global dynamics are thus dictated by Eq. (\ref{evol})
and are, therefore, unitary. That is to say that there is a unitary
operator, $U_{I}\left(t\right),$ acting on the qubit and the boson
bath, such that
\begin{eqnarray}
\rho_{I}\left(t\right) & = & U_{I}\left(t\right)\rho_{I}\left(0\right)U_{I}^{\dagger}\left(t\right),\label{UI}
\end{eqnarray}
with
\begin{eqnarray}
i\hbar\frac{d}{dt}U_{I}\left(t\right) & = & H_{I}\left(t\right)U_{I}\left(t\right)\label{UIevol}
\end{eqnarray}
and
\begin{eqnarray}
U_{I}\left(0\right) & = & \mathbb{I}_{S}\otimes\mathbb{I}_{B},\label{UI(0)}
\end{eqnarray}
where $\mathbb{I}_{B}$ is the unitary operator acting on the boson
Hilbert space. In Eq. (\ref{UI}), the initial global state here is
prepared at $t=0$ as a factored state:
\begin{eqnarray}
\rho_{I}\left(0\right) & = & \rho_{S}\left(0\right)\rho_{B}\left(0\right),\label{initial}
\end{eqnarray}
where $\rho_{S}\left(0\right)$ and $\rho_{B}\left(0\right)$ are
given by Eqs. (\ref{rhoS(0)}) and (\ref{thermal}), respectively.
From Eqs. (\ref{UI}) and (\ref{initial}), it follows that
\begin{eqnarray}
\rho_{IS}\left(t\right) & = & \mathrm{Tr}_{B}\left[U_{I}\left(t\right)\rho_{S}\left(0\right)\rho_{B}\left(0\right)U_{I}^{\dagger}\left(t\right)\right],\label{rhoIS(t)}
\end{eqnarray}
which is the reduced density operator that describes the state of
the logical qubit, also in the interaction picture.

From now on let us adopt the convenient index notation in which $\left(\mathbb{I}_{S},\sigma_{x},\sigma_{y},\sigma_{z}\right)=\left(\sigma_{0},\sigma_{1},\sigma_{2},\sigma_{3}\right).$
We thus can write
\begin{eqnarray}
U_{I}\left(t\right) & = & \sum_{\mu=0}^{3}B_{\mu}\left(t\right)\sigma_{\mu},\label{UI(t)}
\end{eqnarray}
noticing that the operators $B_{\mu}\left(t\right),$ for $\mu=0,1,2,3,$
act only on the boson states. Using Eqs. (\ref{rhoIS(t)}) and (\ref{UI(t)}),
we obtain
\begin{eqnarray}
\rho_{IS}\left(t\right) & = & \sum_{\mu=0}^{3}\sum_{\nu=0}^{3}\sigma_{\mu}\rho_{S}\left(0\right)\sigma_{\nu}M_{\nu,\mu}\left(t\right),\label{rhoIS(t)-1}
\end{eqnarray}
where we have defined the process-matrix elements~\cite{michaelnielsen2011}:
\begin{eqnarray}
M_{\nu,\mu}\left(t\right) & \equiv & \mathrm{Tr}_{B}\left[B_{\nu}^{\dagger}\left(t\right)B_{\mu}\left(t\right)\rho_{B}\left(0\right)\right].\label{Mnumu}
\end{eqnarray}
All we have to do next is to calculate the functions $M_{\nu,\mu}\left(t\right),$
for $\nu,\mu\in\left\{ 0,1,2,3\right\} .$

\section{Numerical simulations}
\label{sec:numer-simul}
Our simulations can be carried out by solving the master equation~\cite{Shibata1977,Chaturvedi1979},\begin{widetext}
\begin{eqnarray}
\frac{d}{dt}\rho_{IS}\left(t\right) & = & -\frac{1}{\hbar^{2}}\mathrm{Tr}_{B}\left\{ \int_{0}^{t}dt^{\prime}\,\left[H_{I}\left(t\right),\left[H_{I}\left(t^{\prime}\right),\rho_{B}\left(0\right)\rho_{IS}\left(t\right)\right]\right]\right\} ,\label{Master}
\end{eqnarray}
or by solving directly the Bloch-vector trajectory equation~\cite{PhysRevA.75.022329},
\begin{eqnarray}
\frac{d}{dt}\mathbf{r}\left(t\right) & = & -4\boldsymbol{\Lambda}\left(t\right)\boldsymbol{\times}\mathrm{Im}\left[\mathbf{G}\left(t\right)\right]-4\boldsymbol{\Lambda}\left(t\right)\boldsymbol{\times}\left\{ \mathbf{r}\left(t\right)\boldsymbol{\times}\mathrm{Re}\left[2\mathbf{F}\left(t\right)+\mathbf{G}\left(t\right)\right]\right\} .\label{trajectory}
\end{eqnarray}
\end{widetext}It is relevant to emphasize that Eq. (\ref{Master})
is time local and it applies to non-Markovian situations, including
the present case, where we assume that the control by continuous dynamical
decoupling is faster than the correlation time of the bath operators~\cite{PhysRevA.75.022329}.
Equation (\ref{trajectory}) is derived from Eq. (\ref{Master}) using
the convenient definitions:
\begin{eqnarray}
\boldsymbol{\Lambda}\left(t\right) & \equiv & \mathbf{\hat{z}}\cos\left[2\Phi\left(t\right)\right]+\mathbf{\hat{y}}\sin\left[2\Phi\left(t\right)\right],\label{Lambda}
\end{eqnarray}
\begin{eqnarray}
\mathbf{F}\left(t\right) & \equiv & \int_{0}^{t}dt^{\prime}\,\boldsymbol{\Lambda}\left(t^{\prime}\right)\mathcal{I}_{1}\left(t-t^{\prime}\right),\label{F(t)}
\end{eqnarray}
\begin{eqnarray}
\mathbf{G}\left(t\right) & \equiv & \int_{0}^{t}dt^{\prime}\,\boldsymbol{\Lambda}\left(t^{\prime}\right)\mathcal{I}_{2}\left(t-t^{\prime}\right),\label{G(t)}
\end{eqnarray}
\begin{eqnarray}
\mathcal{I}_{1}\left(t\right) & \equiv & \sum_{s}\left|g_{s}\right|^{2}\frac{\exp\left(i\omega_{s}t\right)}{\exp\left(\beta\hbar\omega_{s}\right)-1},\label{I1}
\end{eqnarray}
and
\begin{eqnarray}
\mathcal{I}_{2}\left(t\right) & \equiv & \sum_{s}\left|g_{s}\right|^{2}\exp\left(i\omega_{s}t\right).\label{I2}
\end{eqnarray}
The Bloch vector, $\mathbf{r}\left(t\right),$ is related with the
reduced density-matrix operator in the interaction picture, $\rho_{IS}\left(t\right),$
as~\cite{karlblum2012}
\begin{eqnarray}
\rho_{IS}\left(t\right) & = & \frac{1}{2}\mathbb{I}_{S}+\frac{1}{2}\mathbf{r}\left(t\right)\boldsymbol{\cdot}\boldsymbol{\sigma}.\label{Bloch-rho}
\end{eqnarray}
If we write
\begin{eqnarray}
\mathbf{r}\left(t\right) & = & \mathbf{\hat{x}}x_{1}\left(t\right)+\mathbf{\hat{y}}x_{2}\left(t\right)+\mathbf{\hat{z}}x_{3}\left(t\right),\label{r(t)}
\end{eqnarray}
we can calculate the Bloch vector from the density operator by
\begin{eqnarray}
\mathbf{r}\left(t\right) & = & \mathrm{Tr}\left[\boldsymbol{\sigma}\rho_{IS}\left(t\right)\right],\label{correspondence}
\end{eqnarray}
where the trace without a subscript stands for the trace of $2\times2$
matrices. Of course, the initial condition for the Bloch vector is
easily calculated using Eq. (\ref{correspondence}) at $t=0$ and
Eq. (\ref{rhoS(0)}).

For the sake of our emulation of dephasing we simplify the structure
of the noise by assuming an Ohmic spectral density~\cite{gardiner_quantum_2004},
namely,
\begin{eqnarray}
J\left(\omega\right) & \equiv & \sum_{s}\left|g_{s}\right|^{2}\delta\left(\omega-\omega_{s}\right),\label{continuum}
\end{eqnarray}
with
\begin{eqnarray}
J\left(\omega\right) & = & \eta\omega\exp\left(-\frac{\omega}{\omega_{c}}\right).\label{ohmic}
\end{eqnarray}
Here, $\omega_{c}$ is a cutoff frequency and $\eta$ is a dimensionless
noise strength. Using Eqs. (\ref{continuum}) and (\ref{ohmic}) in
Eqs. (\ref{I1}) and (\ref{I2}), we obtain:
\begin{eqnarray}
\mathcal{I}_{1}\left(t\right) & = & \frac{\eta}{\beta^{2}\hbar^{2}}\psi^{\left(1\right)}\left(1+\frac{1}{\beta\omega_{c}\hbar}-\frac{it}{\beta\hbar}\right),\label{I1(t)}
\end{eqnarray}
where $\psi^{\left(1\right)}\left(z\right)$ is the first polygamma
function~\cite{georgearfken2011}, and
\begin{eqnarray}
\mathcal{I}_{2}\left(t\right) & = & \frac{\eta\omega_{c}^{2}}{\left(1-i\omega_{c}t\right)^{2}}.\label{I2(t)}
\end{eqnarray}

\begin{figure}[H]
  \centering
  \includegraphics[width=0.45\textwidth]{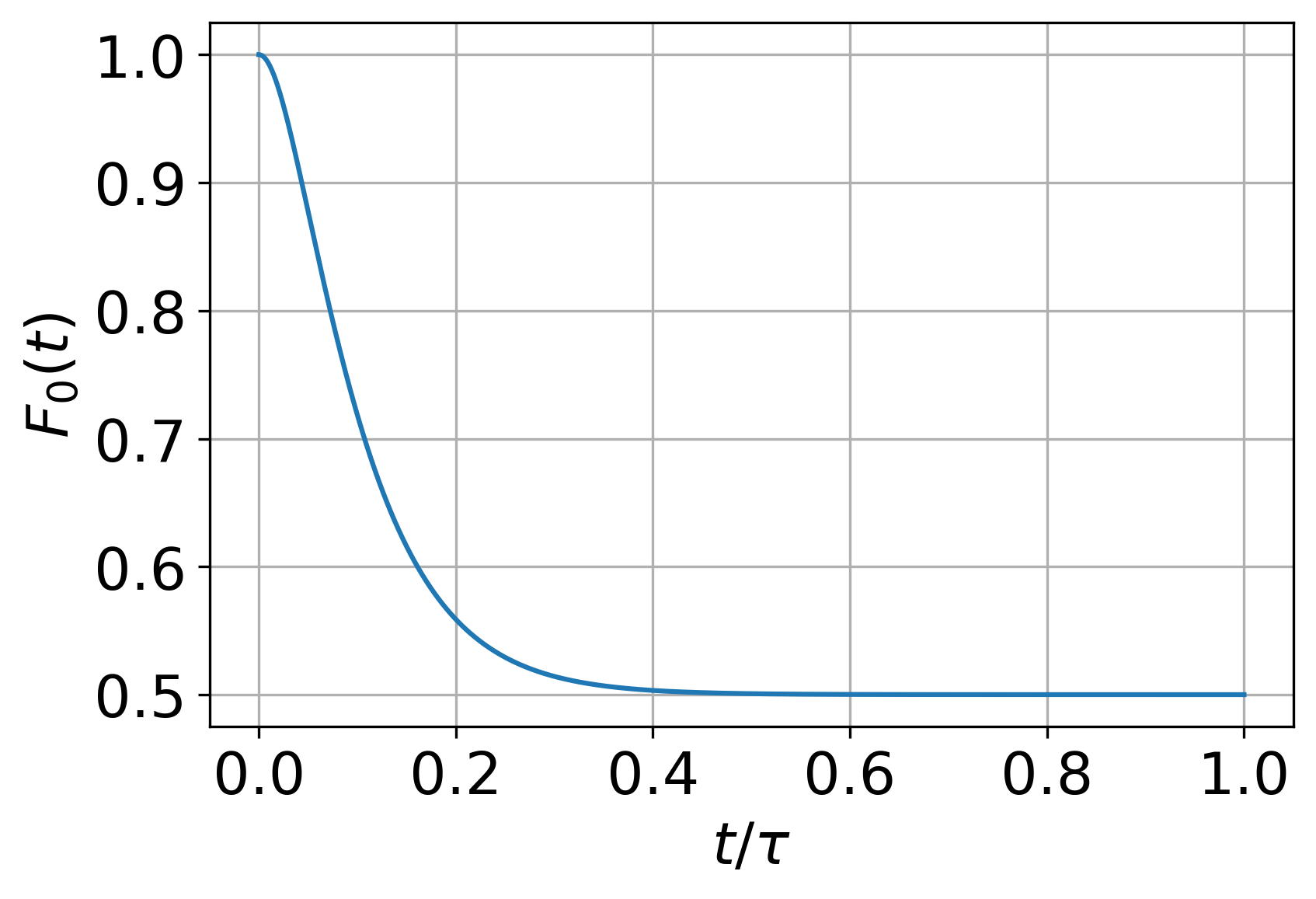}

\caption{\label{pure noise Fig.}The fidelity of Eq. (\ref{pure-noise fidelity})
as a function of time for pure noise. We use $\Omega\left(t\right)=0,$
$\eta=0.01,$ $\omega_{c}=8\pi/\tau,$ $T=\hbar\omega_{c}/k_{B},$
and $c_{1}=ic_{2}=1/\sqrt{2}$ in Eq. (\ref{pure noise}). The final
value of the fidelity, in this case, is $F_{0}\left(\tau\right)\approx0.5296.$}

\end{figure}

To set up our numerical investigations, we establish the numerical
values to characterize the effects of the phase noise. Let us, thus,
look only at the decoherence that the noise produces by choosing $\Omega\left(t\right)=0$
in Eq. (\ref{HG}) and starting with an initially pure density matrix
$\rho_{S}\left(0\right)=\left|\psi_{0}\right\rangle \left\langle \psi_{0}\right|$
derived from the initial qubit state
\begin{eqnarray}
\left|\psi_{0}\right\rangle  & \equiv & c_{1}\left|1\right\rangle +c_{2}\left|2\right\rangle ,\label{pure noise}
\end{eqnarray}
where $c_{1}$ and $c_{2}$ are two complex constants such that
\begin{eqnarray}
\left|c_{1}\right|^{2}+\left|c_{2}\right|^{2} & = & 1.\label{normal c's}
\end{eqnarray}
Since there is no quantum gate action and Eq. (\ref{pure noise}) is an arbitrary coherent state, what we describe here are the effects of the dephasing noise in a quantum memory, where the bath perturbation destroys the coherence and, hence, the quantum information stored in the qubit system. As is derived in Appendix
\ref{=00005Cxi(t)}, the exact analytical solution of the Schr\"{o}dinger
equation using the interaction-picture Hamiltonian of Eq. (\ref{HI}),
for the spectral density of Eqs. (\ref{continuum}) and (\ref{ohmic})
and the initial state of Eq. (\ref{pure noise}), in the absence of
a quantum gate, that is, for $\Omega\left(t\right)=0,$ produces a
reduced density matrix in the interaction picture given by:
\begin{eqnarray}
\rho_{IS}^{\mathrm{pure\,noise}}\left(t\right) & = & \left(\begin{array}{cc}
\left|c_{1}\right|^{2} & c_{1}c_{2}^{\ast}\xi\left(t\right)\\
c_{1}^{\ast}c_{2}\xi\left(t\right) & \left|c_{2}\right|^{2}
\end{array}\right),\label{exact}
\end{eqnarray}
where
\begin{eqnarray}
\xi\left(t\right) & \equiv & \left\{ \frac{\left|\left(\frac{k_{B}T}{\hbar\omega_{c}}+i\frac{k_{B}T}{\hbar}t\right)!\right|^{4}}{\left(1+\omega_{c}^{2}t^{2}\right)\left[\left(\frac{k_{B}T}{\hbar\omega_{c}}\right)!\right]^{4}}\right\} ^{2\eta}.\label{decaying coherence}
\end{eqnarray}
Here, of course, we are using the factorial notation for the gamma
function~\cite{georgearfken2011}:
\begin{eqnarray}
z! & \equiv & \Gamma\left(z+1\right)\nonumber \\
 & = & \int_{0}^{\infty}ds\,\exp\left(-s\right)s^{z},\label{factorian as gamma function}
\end{eqnarray}
for $\mathrm{Re}\left(z\right)>-1.$ In the interaction picture, the
fidelity measure, at an instant $t,$ when we have an ideal pure state
$\rho_{IS}^{\mathrm{ideal}}\left(t\right)=\rho_{S}\left(0\right)$
and a general state $\rho_{IS}\left(t\right),$ is given by~\cite{UHLMANN1976273,doi:10.1080/09500349414552171}:
\begin{eqnarray}
F\left(t\right) & \equiv & \mathrm{Tr}\left[\rho_{IS}\left(t\right)\rho_{S}\left(0\right)\right].\label{fidelity measure}
\end{eqnarray}
Figure \ref{pure noise Fig.} shows 
\begin{eqnarray}
F_{0}\left(t\right) & \equiv & \left\langle \psi_{0}\right|\rho_{IS}^{\mathrm{pure\,noise}}\left(t\right)\left|\psi_{0}\right\rangle ,\label{pure-noise fidelity}
\end{eqnarray}
which results from Eq. (\ref{fidelity measure}) for the case of pure
noise, where we use $\Omega\left(t\right)=0,$ $\eta=0.01,$ $\omega_{c}=8\pi/\tau,$
$T=\hbar\omega_{c}/k_{B},$ and $c_{1}=ic_{2}=1/\sqrt{2}$ in Eq.
(\ref{pure noise}). The final value of the fidelity, in this case,
is
\begin{eqnarray}
F_{0}\left(\tau\right) & \approx & 0.5296.\label{final F0}
\end{eqnarray}
It is important to notice that this exact result agrees perfectly
well with our numerical calculations using the master equation of
Eqs. (\ref{Master}) and (\ref{trajectory}).

\begin{figure}[H]
  \centering
  \includegraphics[width=0.45\textwidth]{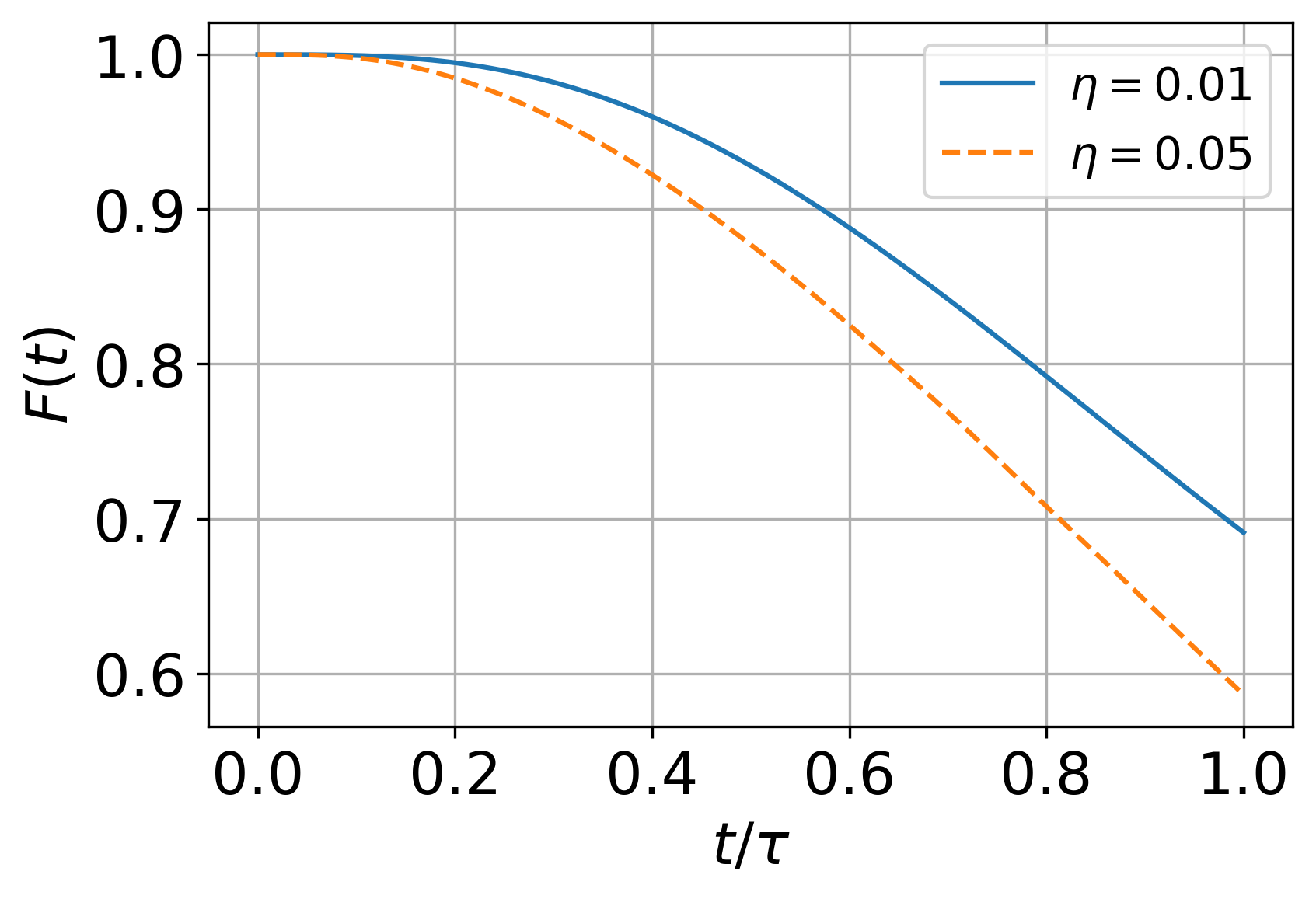}

\caption{\label{=00005COmega n=00003D1} The fidelity of Eq. (\ref{fidelity measure})
as a function of time when we have $\rho_{S}\left(0\right)=\left|1\right\rangle \left\langle 1\right|,$
$\Omega\left(t\right)=\pi/4\tau,$ $\omega_{c}=8\pi/\tau,$ $T=\hbar\omega_{c}/k_{B},$
and two values of the noise strength: $\eta=0.01$ and $\eta=0.05.$
The fidelities we obtain at $t=\tau$ for each of these values of
$\eta$ are, respectively, $F_{\eta=0.01}\left(\tau\right)\approx0.691$
and $F_{\eta=0.05}\left(\tau\right)\approx0.586.$}

\end{figure}

Now that we have compared our numerical simulations against the exact
analytical result, we turn back to the case with $\Omega\left(t\right)\neq0,$
for which we do not have an analytic result and, thus, we use Eqs.
(\ref{Master}) and (\ref{trajectory}) to calculate numerically the
reduced density matrix as a function of time. As mentioned at the
beginning of Sec. \ref{sec:Optimization-of-the}, we take $\left|1\right\rangle $
as the initial state, that is, $\rho_{S}\left(0\right)=\left|1\right\rangle \left\langle 1\right|$
and calculate the final fidelity using Eq. (\ref{fidelity measure})
with $\Omega\left(t\right)=\pi/4\tau,$ $\omega_{c}=8\pi/\tau,$ $T=\hbar\omega_{c}/k_{B},$
and two values of the noise strength: $\eta=0.01$ and $\eta=0.05.$
The fidelities we obtain at $t=\tau$ for each of these values of
$\eta$ are, respectively,
\begin{eqnarray}
F_{\eta=0.01}\left(\tau\right) & \approx & 0.691\label{F0.01}
\end{eqnarray}
and
\begin{eqnarray}
F_{\eta=0.05}\left(\tau\right) & \approx & 0.586.\label{F0.05}
\end{eqnarray}
Figure \ref{=00005COmega n=00003D1} shows the fidelities $F_{\eta=0.01}\left(t\right)$
and $F_{\eta=0.01}\left(t\right)$ as functions of time.

\begin{figure}[H]
  \centering
  \includegraphics[scale=0.9]{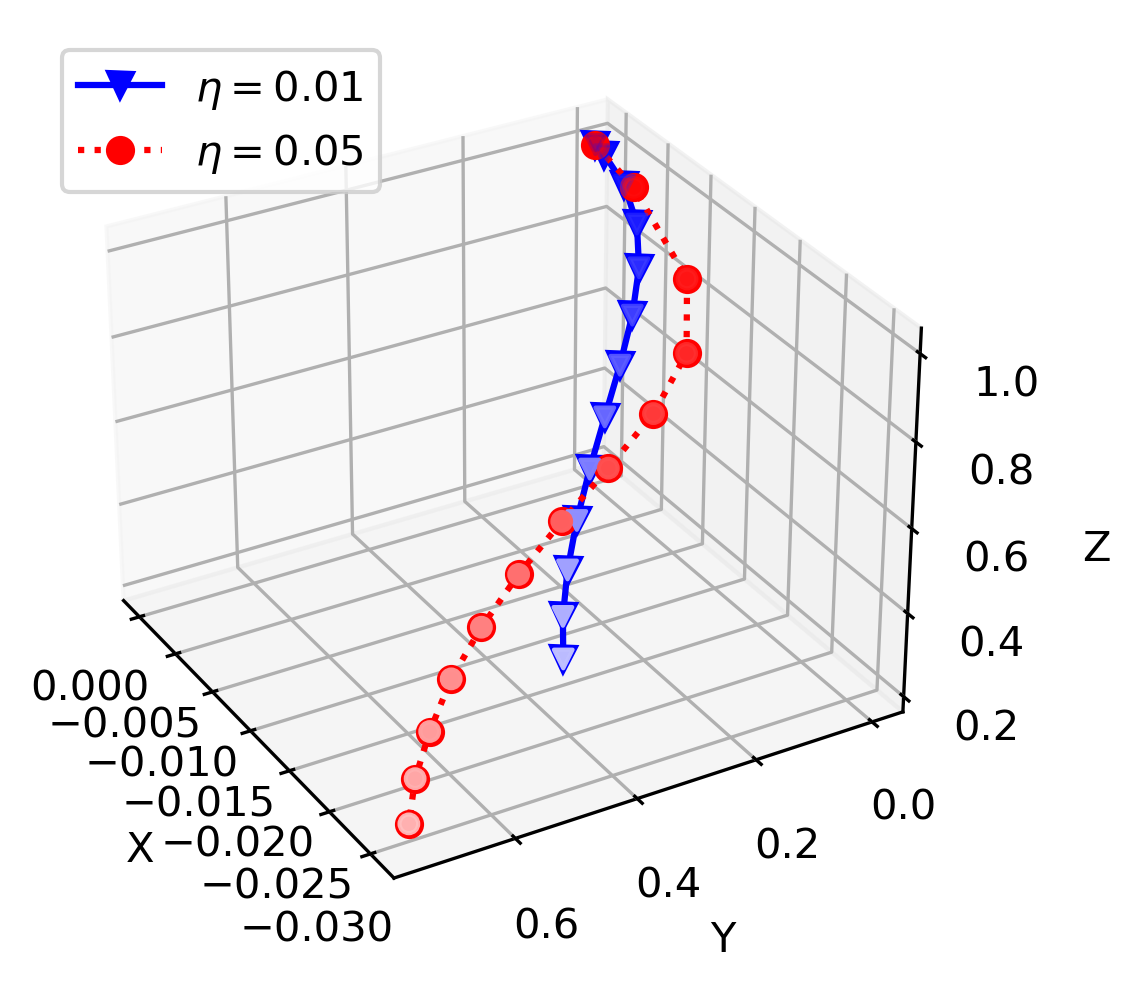}

  \caption{\label{3D trajectory} Bloch-vector trajectories obtained by
    solving Eq. (\ref{trajectory}) when we have
    $\rho_{S}\left(0\right)=\left|1\right\rangle \left\langle 1\right|,$
    $\Omega\left(t\right)=\pi/4\tau,$ $\omega_{c}=8\pi/\tau,$ and
    $T=\hbar\omega_{c}/k_{B},$ for $\eta=0.01$ and $\eta=0.05,$ in which
    cases we obtain the final fidelities of Eqs. (\ref{F0.01}) and
    (\ref{F0.05}), that is, $F_{\eta=0.01}\left(\tau\right)\approx0.691$
    and $F_{\eta=0.05}\left(\tau\right)\approx0.586,$ respectively. Time
    evolution proceeds from darker to lighter. For \(\eta = 0.01\),
    \(\mathbf{r}(\tau) = -0.028\hat{\mathbf{x}}+ 0.45\hat{\mathbf{y}}+
    0.38\hat{\mathbf{z}}\). For \(\eta = 0.05\),
    \(\mathbf{r}(\tau) = -0.03\hat{\mathbf{x}}+ 0.74\hat{\mathbf{y}}+
    0.17\hat{\mathbf{z}}\).}
\end{figure}

We are now able to calculate the trajectories described by the Bloch-vector
evolution calculated using Eq. (\ref{trajectory}). Figure \ref{3D trajectory}
shows, as illustrations, the trajectories for the cases whose final
fidelities are given by Eqs. (\ref{F0.01}) and (\ref{F0.05}).

\begin{figure}[h]
  \centering
  \includegraphics[width=0.45\textwidth]{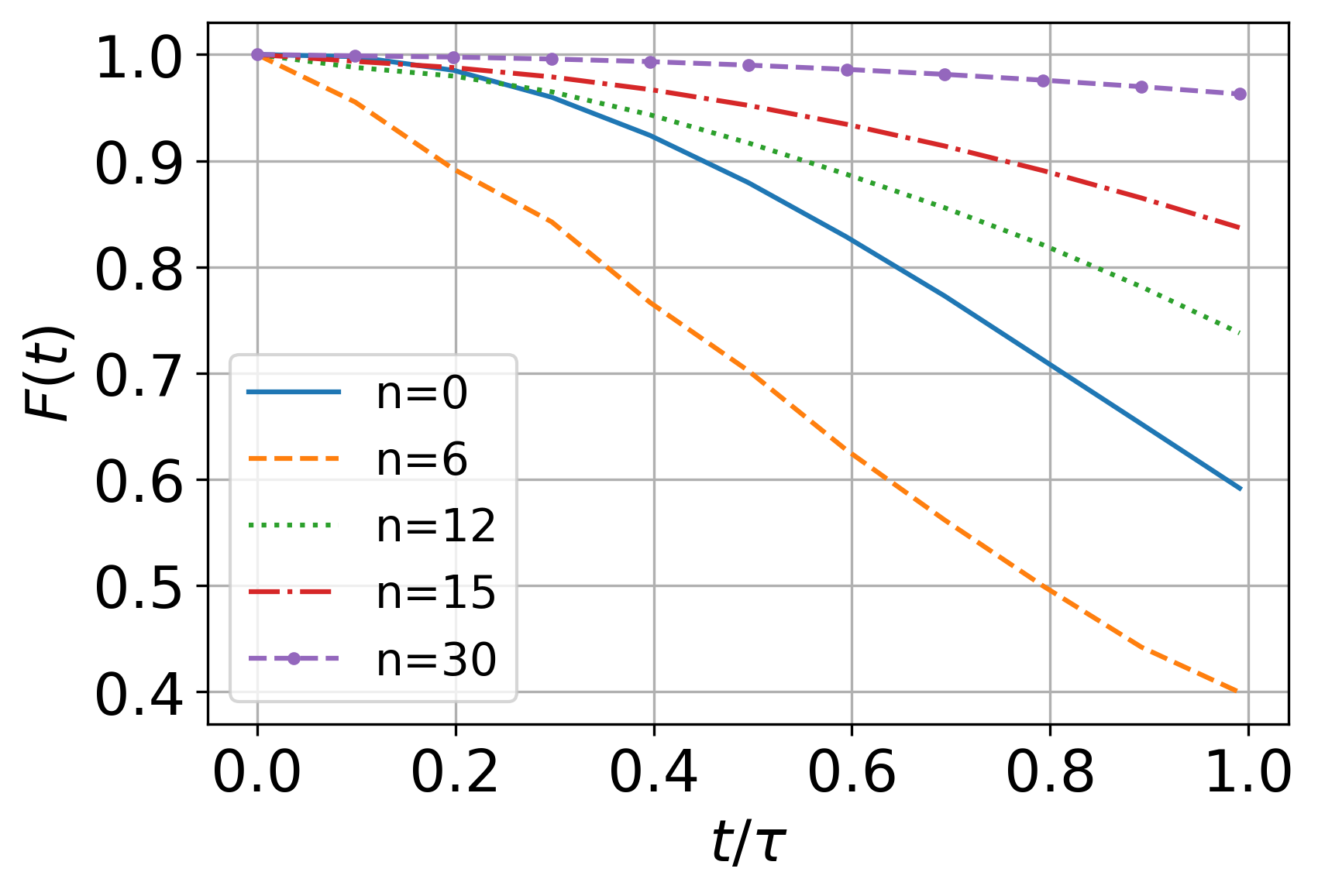} 
\caption{\label{Continuous dynamical decoupling} Fidelity measures as functions
of time, Eq. (\ref{fidelity measure}), using different values of
$n$ appearing in Eq. (\ref{cdd}). Here we have used $\rho_{S}\left(0\right)=\left|1\right\rangle \left\langle 1\right|,$
$\omega_{c}=8\pi/\tau,$ $T=\hbar\omega_{c}/k_{B},$ $\eta=0.05,$
and $\Omega\left(t\right)$ as given by Eq. (\ref{cdd}), for the
cases where $n=0,$ $n=6,$ $n=12,$ $n=15,$ and $n=30.$}

\end{figure}

The optimization procedure we adopt here is called the continuous
dynamical decoupling of phase noise, as prescribed in Ref. ~\cite{PhysRevA.75.022329}.
Accordingly, for protecting against dephasing, in the case of our
Hamiltonian of Eq. (\ref{HS}), we simply have to choose a constant
$\Omega\left(t\right)$ given by
\begin{eqnarray}
\Omega\left(t\right) & = & \frac{\pi}{4\tau}+\frac{2n\pi}{\tau},\label{cdd}
\end{eqnarray}
with $n\in\mathbb{Z}.$ The final fidelity at $t=\tau$ will approach
unity as the magnitude of $n$ is sufficiently increased. Figure \ref{Continuous dynamical decoupling}
shows the fidelity measures as functions of time, Eq. (\ref{fidelity measure}),
using different values of $n$ appearing in Eq. (\ref{cdd}). In the
simulations of Fig. \ref{Continuous dynamical decoupling} we have
used $\rho_{S}\left(0\right)=\left|1\right\rangle \left\langle 1\right|,$
$\omega_{c}=8\pi/\tau,$ $T=\hbar\omega_{c}/k_{B},$ $\eta=0.05,$
and $\Omega\left(t\right)$ as given by Eq. (\ref{cdd}), for the
cases where $n=0,$ $n=6,$ $n=12,$ $n=15,$ and $n=30.$ Table \ref{cddTable}
shows the final fidelity measures, Eq. (\ref{fidelity measure}),
for these choices of $n.$

\begin{table}[h]
\begin{tabular}{|c||c|c|c|c|c|}
\hline 
$n$ in Eq. (\ref{cdd}) & $0$ & $6$ & $12$ & $15$ & $30$\tabularnewline
\hline 
\hline 
$F\left(\tau\right)$ from Eq. (\ref{fidelity measure}) & $0.586$ & $0.391$ & $0.734$ & $0.836$ & $0.962$\tabularnewline
\hline 
$\left|\mathbf{r}\left(\tau\right)-\mathbf{r}\left(0\right)\right|$ & $1.11$ & $1.46$ & $0.97$ & $0.78$ & $0.38$\tabularnewline
\hline 
\end{tabular}

\caption{\label{cddTable} Final fidelities at $t=\tau,$ Eq. (\ref{fidelity measure}),
with $\rho_{S}\left(0\right)=\left|1\right\rangle \left\langle 1\right|,$
$\omega_{c}=8\pi/\tau,$ $T=\hbar\omega_{c}/k_{B},$ $\eta=0.05,$
and $\Omega\left(t\right)$ as given by Eq. (\ref{cdd}), for $n=0,$
$n=6,$ $n=12,$ $n=15,$ and $n=30.$ Also shown is the distance
between the final Bloch vector, $\mathbf{r}\left(\tau\right),$ and
its initial value, $\mathbf{r}\left(0\right)=\mathbf{\hat{z}},$ as
a function of $n$ for the values chosen.}
\end{table}
It is interesting that as $n$ increases the final fidelity initially
decreases and, after $n$ reaches a sufficiently higher magnitude
the fidelity starts to approach unity monotonically. Specifically
for the choice of parameters and initial state of Fig. \ref{Continuous dynamical decoupling},
we observe that the final fidelity progressively decreases to values
lower than the one for $n=0,$ Eq. (\ref{F0.05}), as $n$ is changed
from $n=1$ to $n=5,$ and then increases monotonically with $n\geqslant6,$
surpassing the value given by Eq. (\ref{F0.05}) for $n\geqslant10.$

\begin{figure}
  \centering
  \includegraphics[scale=0.8]{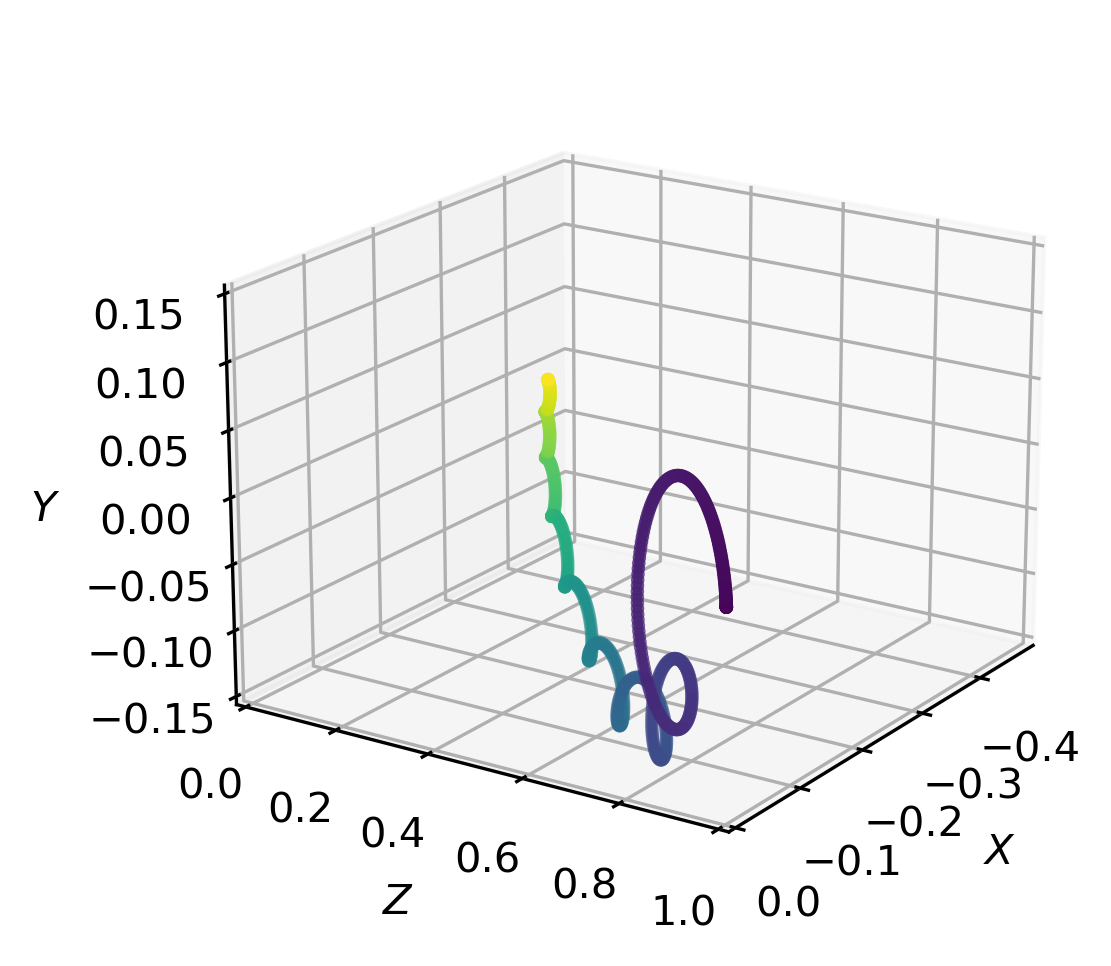}
  \caption{\label{fig-cdd-bloch} The Bloch-vector trajectory when $n=2$
    for
    $\rho_{S}\left(0\right)=\left|1\right\rangle \left\langle 1\right|,$
    $\omega_{c}=8\pi/\tau,$ $T=\hbar\omega_{c}/k_{B},$ $\eta=0.05,$ and
    $\Omega\left(t\right)$ as given by Eq. (\ref{cdd}). Time evolution
    proceeds from purple (darker) to yellow (lighter). It starts at
    \(\mathbf{r}(0) = \hat{\mathbf{z}}\) and ends at
    \(\mathbf{r}(\tau) = -0.485\hat{\mathbf{x}} - 0.024\hat{\mathbf{y}}
    - 0.045\hat{\mathbf{z}}\).}
\end{figure}

As an illustration of a trajectory traced by the evolution of a Bloch
vector under continuous dynamical decoupling, Fig. \ref{fig-cdd-bloch}
shows such a curve corresponding to the case with $n=2$ when $\rho_{S}\left(0\right)=\left|1\right\rangle \left\langle 1\right|,$
$\omega_{c}=8\pi/\tau,$ $T=\hbar\omega_{c}/k_{B},$ $\eta=0.05,$
and $\Omega\left(t\right)$ as given by Eq. (\ref{cdd}). In the interaction
picture, if $n$ is large enough, the distance between $\mathbf{r}\left(\tau\right)$
and $\mathbf{r}\left(0\right)$ gradually decreases, as is also shown
in Table \ref{cddTable}. Moreover, as $n$ increases, the frequency
$\Omega\left(t\right),$ Eq. (\ref{cdd}), gets large and the helical
aspect of the Bloch-vector trajectories becomes progressively more
pronounced than the one shown in Fig. \ref{fig-cdd-bloch} for $n=2.$

Once we obtain the Bloch-vector trajectory, $\mathbf{r}\left(t\right),$
or, equivalently, the density matrix, $\rho_{IS}\left(t\right),$
as functions of time, we must calculate the functions given by Eq.
(\ref{Mnumu}). We have proceeded as follows. From Eq. (\ref{rhoIS(t)-1})
we see that the matrix elements $M_{\nu,\mu}\left(t\right)$ do not
depend on the initial condition. Therefore, we can choose different
initial conditions to obtain a sufficient number of equations that
we can then solve to obtain all these elements. Accordingly, we have
chosen the following four different initial conditions for the components
of the Bloch vector:
\begin{eqnarray}
x_{k,\mu}\left(0\right) & = & \delta_{k,\mu},\label{xkmu}
\end{eqnarray}
for $k=1,2,3$ and $\mu=0,1,2,3.$ Now, after time $t>0,$ we will
have four evolved Bloch vectors given by
\begin{eqnarray}
\mathbf{r}_{\mu}\left(t\right) & \equiv & \mathbf{\hat{x}}x_{1,\mu}\left(t\right)+\mathbf{\hat{y}}x_{2,\mu}\left(t\right)+\mathbf{\hat{z}}x_{3,\mu}\left(t\right),\label{rmu(0)}
\end{eqnarray}
for $\mu=0,1,2,3.$ Using the initial conditions of Eq. (\ref{xkmu}),
we obtain the following equations to solve for the process-matrix elements
$M_{\nu,\mu}\left(t\right):$\begin{widetext}
\begin{eqnarray}
\sum_{\alpha=0}^{3}\sum_{\beta=0}^{3}\sigma_{\alpha}\left[\frac{1}{2}\mathbb{I}_{S}+\frac{1}{2}\mathbf{r}_{\mu}\left(0\right)\boldsymbol{\cdot}\boldsymbol{\sigma}\right]\sigma_{\beta}M_{\beta,\alpha}\left(t\right) & = & \frac{1}{2}\mathbb{I}_{S}+\frac{1}{2}\mathbf{r}_{\mu}\left(t\right)\boldsymbol{\cdot}\boldsymbol{\sigma},\label{M-eqs}
\end{eqnarray}
\end{widetext}for $\mu=0,1,2,3,$ where we have used Eqs. (\ref{rhoIS(t)-1})
and (\ref{Bloch-rho}), together with the initial conditions given
by Eq. (\ref{xkmu}). Here, of course, $\mathbf{r}_{\mu}\left(t\right)$
are the evolved Bloch vectors obtained using the corresponding initial
conditions given by Eq. (\ref{rmu(0)}). By solving Eq. (\ref{M-eqs})
explicitly, we obtain:\begin{eqnarray}
M_{0,0}&=& \frac{1}{4} \left(-x_{1,0}+x_{1,1}-x_{2,0}+x_{2,2}-x_{3,0}+x_{3,3}+1\right),\nonumber\\
&&\\
M_{0,1}&=& \frac{1}{4} \left(x_{1,0}-i x_{2,0}+i x_{2,3}+i x_{3,0}-i x_{3,2}\right),\\
M_{0,2}&=& \frac{1}{4} i \left(x_{1,0}-x_{1,3}-i x_{2,0}-x_{3,0}+x_{3,1}\right),\\
M_{0,3}&=& -\frac{1}{4} i \left(x_{1,0}-x_{1,2}-x_{2,0}+x_{2,1}+i x_{3,0}\right),\\
M_{1,0}&=& \frac{1}{4} \left(x_{1,0}+i x_{2,0}-i x_{2,3}-i x_{3,0}+i x_{3,2}\right),\\
M_{1,1}&=& \frac{1}{4} \left(-x_{1,0}+x_{1,1}+x_{2,0}-x_{2,2}+x_{3,0}-x_{3,3}+1\right),\nonumber\\
&&\\
M_{1,2}&=& \frac{1}{4} \left(-x_{1,0}+x_{1,2}-x_{2,0}+x_{2,1}+i x_{3,0}\right),\\
M_{1,3}&=& \frac{1}{4} \left(-x_{1,0}+x_{1,3}-i x_{2,0}-x_{3,0}+x_{3,1}\right),\\
M_{2,0}&=& -\frac{1}{4} i \left(x_{1,0}-x_{1,3}+i x_{2,0}-x_{3,0}+x_{3,1}\right),\\
M_{2,1}&=& \frac{1}{4} \left(-x_{1,0}+x_{1,2}-x_{2,0}+x_{2,1}-i x_{3,0}\right),\\
M_{2,2}&=& \frac{1}{4} \left(x_{1,0}-x_{1,1}-x_{2,0}+x_{2,2}+x_{3,0}-x_{3,3}+1\right),\nonumber\\
&&\\
M_{2,3}&=& \frac{1}{4} i \left(x_{1,0}+i x_{2,0}-i x_{2,3}+i x_{3,0}-i x_{3,2}\right),\\
M_{3,0}&=& \frac{1}{4} i \left(x_{1,0}-x_{1,2}-x_{2,0}+x_{2,1}-i x_{3,0}\right),\\
M_{3,1}&=& \frac{1}{4} \left(-x_{1,0}+x_{1,3}+i x_{2,0}-x_{3,0}+x_{3,1}\right),\\
M_{3,2}&=& -\frac{1}{4} i \left(x_{1,0}-i x_{2,0}+i x_{2,3}-i x_{3,0}+i x_{3,2}\right),\\
M_{3,3}&=& \frac{1}{4} \left(x_{1,0}-x_{1,1}+x_{2,0}-x_{2,2}-x_{3,0}+x_{3,3}+1\right),\nonumber\\
&&
\end{eqnarray}where, we have simplified the notation by writing $M_{\nu,\mu}=M_{\nu,\mu}\left(t\right)$
and $x_{k,\mu}=x_{k,\mu}\left(t\right),$ with $k\in\left\{ 1,2,3\right\} $
and $\nu,\mu\in\left\{ 0,1,2,3\right\} .$

\section{Visualization of Kraus operators}
\label{sec:visu-kraus-oper}

As can be seen from Eq. (\ref{Mnumu}), the matrix whose elements
are the functions $M_{\nu,\mu}\left(t\right)$ is Hermitian. Therefore,
it can be diagonalized as
\begin{eqnarray}
\sum_{\gamma=0}^{3}W_{\gamma,\mu}^{\ast}\left(t\right)D_{\gamma}\left(t\right)W_{\gamma,\nu}\left(t\right) & = & M_{\mu,\nu}\left(t\right),\label{diag}
\end{eqnarray}
where the matrix whose elements are $W_{\gamma,\nu}\left(t\right)$
is unitary and $D_{\gamma}\left(t\right),$ for $\gamma=0,1,2,3,$
can be easily shown to be non-negative real numbers. The Kraus operators~\cite{kraus1983,KEYL2002431}
are, therefore, defined as~\cite{michaelnielsen2011}
\begin{eqnarray}
K_{\gamma}\left(t\right) & \equiv & \exp\left[i\chi_{\gamma}\left(t\right)\right]\sum_{\mu=0}^{3}\sqrt{D_{\gamma}\left(t\right)}W_{\gamma,\mu}\left(t\right)\sigma_{\mu}\label{K}
\end{eqnarray}
and
\begin{eqnarray}
K_{\gamma}^{\dagger}\left(t\right) & \equiv & \exp\left[-i\chi_{\gamma}\left(t\right)\right]\sum_{\mu=0}^{3}\sqrt{D_{\gamma}\left(t\right)}W_{\gamma,\mu}^{\ast}\left(t\right)\sigma_{\mu},\label{Kdagger}
\end{eqnarray}
where $\chi_{\gamma}\left(t\right),$ for $\gamma=0,1,2,3,$ are arbitrary
real functions, since each Kraus operator can be defined up to a global
phase factor. We are going to make good use of this freedom in what
follows. Next we show, for the present case of the $\sqrt{\mathrm{SWAP}}$
gate under dephasing, how to visualize the evolution of $K_{\gamma}\left(t\right).$

Equations (\ref{HI}), (\ref{sigma(t)}), (\ref{UIevol}), and (\ref{UI(t)})
imply the following coupled matrix equations giving the dynamics of
the boson field operators $B_{\mu}\left(t\right),$ for $\mu=0,1,2,3:$\begin{widetext}
\begin{eqnarray}
i\hbar\frac{d}{dt}\left[\begin{array}{c}
B_{0}\left(t\right)\\
B_{1}\left(t\right)
\end{array}\right] & = & \left[\begin{array}{cc}
\sin\left[2\Phi\left(t\right)\right] & \cos\left[2\Phi\left(t\right)\right]\\
-i\cos\left[2\Phi\left(t\right)\right] & i\sin\left[2\Phi\left(t\right)\right]
\end{array}\right]H_{BI}\left(t\right)\left[\begin{array}{c}
B_{2}\left(t\right)\\
B_{3}\left(t\right)
\end{array}\right]\label{B0B1}
\end{eqnarray}
and
\begin{eqnarray}
i\hbar\frac{d}{dt}\left[\begin{array}{c}
B_{2}\left(t\right)\\
B_{3}\left(t\right)
\end{array}\right] & = & \left[\begin{array}{cc}
\sin\left[2\Phi\left(t\right)\right] & i\cos\left[2\Phi\left(t\right)\right]\\
\cos\left[2\Phi\left(t\right)\right] & -i\sin\left[2\Phi\left(t\right)\right]
\end{array}\right]H_{BI}\left(t\right)\left[\begin{array}{c}
B_{0}\left(t\right)\\
B_{1}\left(t\right)
\end{array}\right],\label{B2B3}
\end{eqnarray}
\end{widetext}where we have defined the boson Hamiltonian $H_{BI}\left(t\right)$
in the interaction picture as
\begin{eqnarray}
H_{BI}\left(t\right) & \equiv & \hbar\sum_{s}\left[g_{s}b_{s}\exp\left(-i\omega_{s}t\right)+g_{s}^{\ast}b_{s}^{\dagger}\exp\left(i\omega_{s}t\right)\right].\nonumber \\
\label{HBI}
\end{eqnarray}
Given the initial condition for $U_{I}\left(t\right),$ Eq. (\ref{UI(0)}),
we obtain, according with Eq. (\ref{UI(t)}), the initial conditions
for the field operators, namely,
\begin{eqnarray}
\left[\begin{array}{c}
B_{0}\left(0\right)\\
B_{1}\left(0\right)\\
B_{2}\left(0\right)\\
B_{3}\left(0\right)
\end{array}\right] & = & \left[\begin{array}{c}
1\\
0\\
0\\
0
\end{array}\right].\label{initial-B's}
\end{eqnarray}
Hence, formal integration of Eq. (\ref{B2B3}) from $t=0$ to $t^{\prime}$
and, after using Eq. (\ref{initial-B's}), substituting the resulting
expression into Eq. (\ref{B0B1}), gives
\begin{eqnarray}
i\hbar\frac{d}{dt}\left[\begin{array}{c}
B_{0}\left(t\right)\\
B_{1}\left(t\right)
\end{array}\right] & = & \frac{1}{i\hbar}\int_{0}^{t}dt^{\prime}\,\left[\begin{array}{cc}
\cos\left[\zeta\left(t,t^{\prime}\right)\right] & i\sin\left[\zeta\left(t,t^{\prime}\right)\right]\\
i\sin\left[\zeta\left(t,t^{\prime}\right)\right] & \cos\left[\zeta\left(t,t^{\prime}\right)\right]
\end{array}\right]\nonumber \\
 &  & \times H_{BI}\left(t\right)H_{BI}\left(t^{\prime}\right)\left[\begin{array}{c}
B_{0}\left(t^{\prime}\right)\\
B_{1}\left(t^{\prime}\right)
\end{array}\right],\label{second-order-evol}
\end{eqnarray}
where
\begin{eqnarray}
\zeta\left(t,t^{\prime}\right) & \equiv & 2\Phi\left(t\right)-2\Phi\left(t^{\prime}\right).\label{zeta}
\end{eqnarray}
Equation (\ref{second-order-evol}) is easily diagonalized by the time-independent Hermitian
and unitary matrix
\begin{eqnarray}
S & \equiv & \frac{1}{\sqrt{2}}\left[\begin{array}{cc}
1 & 1\\
1 & -1
\end{array}\right].\label{Hermitian-and-unitary}
\end{eqnarray}
By defining the $c\text{-number}$ function
\begin{eqnarray}
h\left(t\right) & \equiv & -\sum_{s}\left|g_{s}\right|^{2}2i\int_{0}^{t}dt^{\prime}\,\int_{0}^{t^{\prime}}dt^{\prime\prime}\,\exp\left[-i\zeta\left(t^{\prime},t^{\prime\prime}\right)\right]\nonumber \\
 &  & \times\sin\left[\omega_{s}\left(t^{\prime\prime}-t^{\prime}\right)\right],\label{h(t)}
\end{eqnarray}
we can express the solution of Eq. (\ref{second-order-evol}) as
\begin{eqnarray}
\left[\begin{array}{c}
B_{0}\left(t\right)\\
B_{1}\left(t\right)
\end{array}\right] & = & \frac{1}{2}\left[\begin{array}{c}
\Gamma\left(t\right)+\exp\left[h\left(t\right)\right]\left[\Gamma\left(t\right)\right]^{\dagger}\\
\Gamma\left(t\right)-\exp\left[h\left(t\right)\right]\left[\Gamma\left(t\right)\right]^{\dagger}
\end{array}\right],\label{B0B1-solution}
\end{eqnarray}
where the operator boson field $\Gamma\left(t\right)$ satisfies the
integro-differential equation:
\begin{eqnarray}
\frac{d}{dt}\Gamma\left(t\right) & = & -\frac{1}{\hbar^{2}}\int_{0}^{t}dt^{\prime}\,\exp\left[i\zeta\left(t,t^{\prime}\right)\right]\nonumber \\
 &  & \times H_{BI}\left(t\right)H_{BI}\left(t^{\prime}\right)\Gamma\left(t^{\prime}\right),\label{Gamma}
\end{eqnarray}
with
\begin{eqnarray}
\Gamma\left(0\right) & = & \mathbb{I}_{B}.\label{Gamma(0)}
\end{eqnarray}
Equation (\ref{Gamma}) is solved iteratively, as in the usual
time-dependent perturbation theory.

Next we observe that Eq. (\ref{Gamma}) shows that $\Gamma\left(t\right)$
results in an operator consisting of a series, each term of which
contains an even number of creation $\left(b_{s}^{\dagger}\right)$
and/or annihilation operators $\left(b_{s}\right)$ as factors. This
means that, according with Eq. (\ref{B0B1-solution}), the operators
$B_{0}\left(t\right)$ and $B_{1}\left(t\right)$ will also be series
of terms containing only an even number of boson-operator factors.
However, as we see from Eq. (\ref{B2B3}), because of the extra Hamiltonian
factor, $B_{2}\left(t\right)$ and $B_{3}\left(t\right)$ will be
operator series whose terms each contains an odd number of boson-operator
factors. Therefore, it follows from Eqs. (\ref{thermal}) and (\ref{Mnumu})
that
\begin{eqnarray}
M_{0,2}\left(t\right) & = & M_{2,0}\left(t\right)=0,\label{M02}
\end{eqnarray}
\begin{eqnarray}
M_{0,3}\left(t\right) & = & M_{3,0}\left(t\right)=0,\label{M03}
\end{eqnarray}
\begin{eqnarray}
M_{1,2}\left(t\right) & = & M_{2,1}\left(t\right)=0,\label{M12}
\end{eqnarray}
and
\begin{eqnarray}
M_{1,3}\left(t\right) & = & M_{3,1}\left(t\right)=0.\label{M13}
\end{eqnarray}
We can, thus, arrange the elements $M_{\nu,\mu}\left(t\right),$ for
$\nu,\mu\in\left\{ 0,1,2,3\right\} ,$ as a block-diagonal process matrix~\cite{michaelnielsen2011}:
\begin{eqnarray}
M\left(t\right) & = & \left[\begin{array}{cccc}
M_{0,0}\left(t\right) & M_{0,1}\left(t\right) & 0 & 0\\
M_{0,1}^{\ast}\left(t\right) & M_{1,1}\left(t\right) & 0 & 0\\
0 & 0 & M_{2,2}\left(t\right) & M_{2,3}\left(t\right)\\
0 & 0 & M_{2,3}^{\ast}\left(t\right) & M_{3,3}\left(t\right)
\end{array}\right].\nonumber \\
\label{M(t)}
\end{eqnarray}
Diagonalization of Eq. (\ref{M(t)}) is analytical in terms of the
elements $M_{\nu,\mu}\left(t\right),$ for $\nu,\mu\in\left\{ 0,1,2,3\right\} .$
Each of the two $2\times2$ blocks, for each instant $t,$ is of the
form
\begin{eqnarray}
A & = & \left[\begin{array}{cc}
c & a+bi\\
a-bi & d
\end{array}\right],\label{matrixA}
\end{eqnarray}
where $a,b,c,d\in\mathbb{R}.$ To simplify the notation, let us write
\begin{eqnarray}
a+bi & \equiv & \sqrt{a^{2}+b^{2}}\exp\left(i\varphi\right)\nonumber \\
 & \equiv & \rho\exp\left(i\varphi\right)\label{a+bi}
\end{eqnarray}
and
\begin{eqnarray}
\delta & \equiv & \frac{c-d}{2}.\label{delta}
\end{eqnarray}
We can, thus, by the usual textbook methods, obtain the diagonalizing
matrix $W\left(t\right),$ whose elements are $W_{\gamma,\nu}\left(t\right),$
for $\gamma,\nu\in\left\{ 0,1,2,3\right\} ,$ appearing in Eq. (\ref{diag}):
\begin{eqnarray}
W\left(t\right) & = & \left[\begin{array}{cccc}
u_{+}^{\ast} & u_{-}^{\ast} & 0 & 0\\
u_{-} & -u_{+} & 0 & 0\\
0 & 0 & \tilde{u}_{+}^{\ast} & \tilde{u}_{-}^{\ast}\\
0 & 0 & \tilde{u}_{-} & -\tilde{u}_{+}
\end{array}\right],\label{W(t)}
\end{eqnarray}
where
\begin{eqnarray}
\left[\begin{array}{c}
u_{+}\\
u_{-}
\end{array}\right] & = & \frac{1}{\sqrt{2}}\left[\begin{array}{c}
\exp\left(i\frac{\varphi}{2}\right)\sqrt{1+\frac{\delta}{\sqrt{\delta^{2}+\rho^{2}}}}\\
\exp\left(-i\frac{\varphi}{2}\right)\sqrt{1-\frac{\delta}{\sqrt{\delta^{2}+\rho^{2}}}}
\end{array}\right]\label{eigenvector}
\end{eqnarray}
and a completely analogous expression for the block involving $\tilde{u}_{+}$
and $\tilde{u}_{-}.$ Of course, since Eq. (\ref{matrixA}) is only
a simplified notation for each time-dependent block of Eq. (\ref{M(t)}),
all the non-zero elements of $W\left(t\right)$ are also functions
of $t.$

After diagonalizing $M\left(t\right),$ we obtain the elements $D_{\gamma}\left(t\right),$
for $\gamma=0,1,2,3,$ appearing in Eq. (\ref{diag}):
\begin{eqnarray}
D_{0}\left(t\right) & \equiv & \frac{c+d}{2}+\sqrt{\delta^{2}+\rho^{2}},\label{D0}
\end{eqnarray}
\begin{eqnarray}
D_{1}\left(t\right) & \equiv & \frac{c+d}{2}-\sqrt{\delta^{2}+\rho^{2}},\label{D1}
\end{eqnarray}
\begin{eqnarray}
D_{2}\left(t\right) & \equiv & \frac{\tilde{c}+\tilde{d}}{2}+\sqrt{\tilde{\delta}^{2}+\tilde{\rho}^{2}},\label{D2}
\end{eqnarray}
and
\begin{eqnarray}
D_{3}\left(t\right) & \equiv & \frac{\tilde{c}+\tilde{d}}{2}-\sqrt{\tilde{\delta}^{2}+\tilde{\rho}^{2}},\label{D3}
\end{eqnarray}
where, as above, we use tildes to indicate the corresponding quantities
in the lower $2\times2$ diagonal block of the matrix appearing in
Eq. (\ref{M(t)}). Using Eqs. (\ref{W(t)}) and (\ref{eigenvector})
in the definition of the Kraus operators, Eq. (\ref{K}), we obtain:
\begin{eqnarray}
K_{0}\left(t\right) & \equiv & \sqrt{D_{0}\left(t\right)}\exp\left(i\frac{\varphi}{2}\right)\left(u_{+}^{\ast}\mathbb{I}+u_{-}^{\ast}\sigma_{x}\right)\nonumber \\
 & = & \sqrt{D_{0}\left(t\right)}\left[\left|u_{+}\right|\mathbb{I}+\exp\left(i\varphi\right)\left|u_{-}\right|\sigma_{x}\right],\label{K0}
\end{eqnarray}
\begin{eqnarray}
K_{1}\left(t\right) & \equiv & \sqrt{D_{1}\left(t\right)}\exp\left(i\frac{\varphi}{2}\right)\left(u_{-}\mathbb{I}-u_{+}\sigma_{x}\right)\nonumber \\
 & = & \sqrt{D_{1}\left(t\right)}\left[\left|u_{-}\right|\mathbb{I}-\exp\left(i\varphi\right)\left|u_{+}\right|\sigma_{x}\right],\label{K1}
\end{eqnarray}
\begin{eqnarray}
K_{2}\left(t\right) & \equiv & \sqrt{D_{2}\left(t\right)}\exp\left(-i\frac{\tilde{\varphi}}{2}\right)\left(\tilde{u}_{+}^{\ast}\sigma_{y}+\tilde{u}_{-}^{\ast}\sigma_{z}\right)\nonumber \\
 & = & \sigma_{z}\sqrt{D_{2}\left(t\right)}\left[\left|\tilde{u}_{-}\right|\mathbb{I}-i\exp\left(-i\tilde{\varphi}\right)\left|\tilde{u}_{+}\right|\sigma_{x}\right],\nonumber \\
 \label{K2}
\end{eqnarray}
and
\begin{eqnarray}
K_{3}\left(t\right) & \equiv & -\sqrt{D_{3}\left(t\right)}\exp\left(-i\frac{\tilde{\varphi}}{2}\right)\left(\tilde{u}_{-}\sigma_{y}-\tilde{u}_{+}\sigma_{z}\right)\nonumber \\
 & = & \sigma_{z}\sqrt{D_{3}\left(t\right)}\left[\left|\tilde{u}_{+}\right|\mathbb{I}+i\exp\left(-i\tilde{\varphi}\right)\left|\tilde{u}_{-}\right|\sigma_{x}\right].\nonumber \\
 \label{K3}
\end{eqnarray}
Notice that, as we have already mentioned just following Eqs. (\ref{K})
and (\ref{Kdagger}), we have chosen phase factors whose convenience
we clarify below.

Because $W\left(t\right),$ Eq. (\ref{W(t)}), is a unitary matrix,
we have
\begin{eqnarray}
\left|u_{+}\right|^{2}+\left|u_{-}\right|^{2} & = & 1\label{normalization}
\end{eqnarray}
and
\begin{eqnarray}
\left|\tilde{u}_{+}\right|^{2}+\left|\tilde{u}_{-}\right|^{2} & = & 1,\label{normalization-tilde}
\end{eqnarray}
so that each of the Kraus operators as defined in Eqs. (\ref{K0}-\ref{K3})
can be characterized by only three independent real numbers. Now let
us apply each of these Kraus operators to the ket $\left|1\right\rangle $
of the qubit subspace we are considering, spanned by the basis set
established in Eqs. (\ref{sigmaz1}) and (\ref{sigmaz2}), that is,
$\left\{ \left|1\right\rangle ,\left|2\right\rangle \right\} .$ Of
course, this choice is arbitrary, but it is the analogous choice in
the conventional Bloch-vector representation of qubit density operators~\cite{michaelnielsen2011}.
Because of Eqs. (\ref{sigmax1}) and (\ref{sigmax2}), we obtain:
\begin{eqnarray}
K_{\gamma}\left(t\right)\left|1\right\rangle  & = & \sqrt{D_{\gamma}\left(t\right)}\left|\phi_{\gamma}\right\rangle ,\label{Kgamma}
\end{eqnarray}
for $\gamma=0,1,2,3,$ where
\begin{eqnarray}
\left|\phi_{0}\right\rangle  & = & \left|u_{+}\right|\left|1\right\rangle +\exp\left(i\varphi\right)\left|u_{-}\right|\left|2\right\rangle ,\label{Phi0}
\end{eqnarray}
\begin{eqnarray}
\left|\phi_{1}\right\rangle  & = & \left|u_{-}\right|\left|1\right\rangle -\exp\left(i\varphi\right)\left|u_{+}\right|\left|2\right\rangle ,\label{Phi1}
\end{eqnarray}
\begin{eqnarray}
\left|\phi_{2}\right\rangle  & = & \left|\tilde{u}_{-}\right|\left|1\right\rangle +i\exp\left(-i\tilde{\varphi}\right)\left|\tilde{u}_{+}\right|\left|2\right\rangle ,\label{Phi2}
\end{eqnarray}
and
\begin{eqnarray}
\left|\phi_{3}\right\rangle  & = & \left|\tilde{u}_{+}\right|\left|1\right\rangle -i\exp\left(-i\tilde{\varphi}\right)\left|\tilde{u}_{-}\right|\left|2\right\rangle .\label{Phi3}
\end{eqnarray}
Since Eqs. (\ref{normalization}) and (\ref{normalization-tilde})
hold, these four kets are normalized to unity and each one of them
has a corresponding Bloch vector. Thus, for each ket of Eqs. (\ref{Phi0}-\ref{Phi3})
we define, respectively,
\begin{eqnarray}
\left[\begin{array}{c}
\cos\left(\frac{\theta_{0}}{2}\right)\\
\sin\left(\frac{\theta_{0}}{2}\right)\exp\left(i\varphi_{0}\right)
\end{array}\right] & \equiv & \left[\begin{array}{c}
\left|u_{+}\right|\\
\left|u_{-}\right|\exp\left(i\varphi\right)
\end{array}\right],\label{angles0}
\end{eqnarray}
\begin{eqnarray}
\left[\begin{array}{c}
\cos\left(\frac{\theta_{1}}{2}\right)\\
\sin\left(\frac{\theta_{1}}{2}\right)\exp\left(i\varphi_{1}\right)
\end{array}\right] & \equiv & \left[\begin{array}{c}
\left|u_{-}\right|\\
-\left|u_{+}\right|\exp\left(i\varphi\right)
\end{array}\right],\label{angles1}
\end{eqnarray}
\begin{eqnarray}
\left[\begin{array}{c}
\cos\left(\frac{\theta_{2}}{2}\right)\\
\sin\left(\frac{\theta_{2}}{2}\right)\exp\left(i\varphi_{2}\right)
\end{array}\right] & \equiv & \left[\begin{array}{c}
\left|\tilde{u}_{-}\right|\\
i\exp\left(-i\tilde{\varphi}\right)\left|\tilde{u}_{+}\right|
\end{array}\right],\label{angles2}
\end{eqnarray}
and
\begin{eqnarray}
\left[\begin{array}{c}
\cos\left(\frac{\theta_{3}}{2}\right)\\
\sin\left(\frac{\theta_{3}}{2}\right)\exp\left(i\varphi_{3}\right)
\end{array}\right] & \equiv & \left[\begin{array}{c}
\left|\tilde{u}_{+}\right|\\
-i\exp\left(-i\tilde{\varphi}\right)\left|\tilde{u}_{-}\right|
\end{array}\right].\label{angles3}
\end{eqnarray}

\begin{figure}[H]
  \centering
  \subfloat[]{
    \includegraphics[scale=0.45]{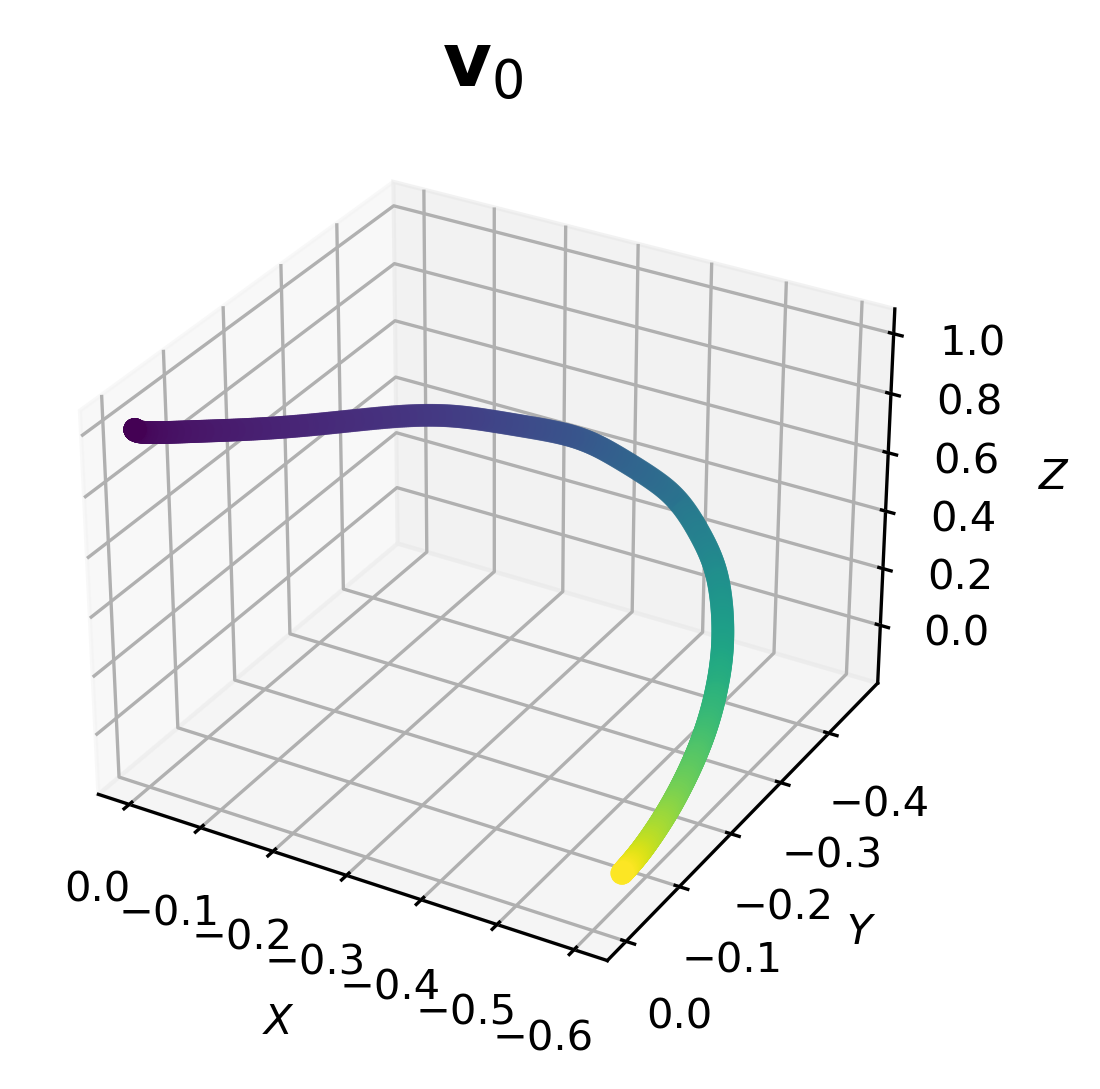}
  }
  \subfloat[]{
    \includegraphics[scale=0.45]{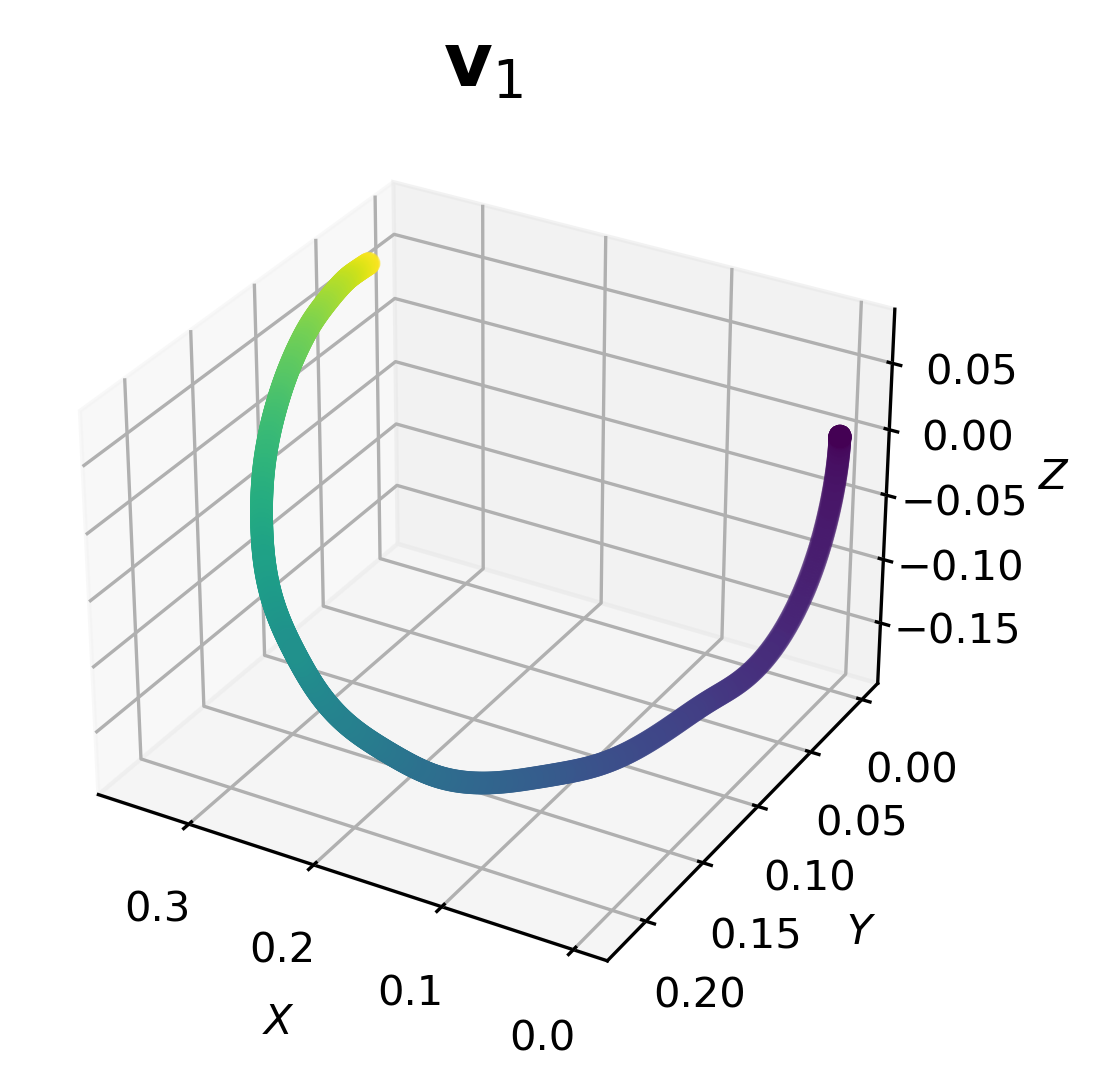}
  }
  \newline
    \subfloat[]{
    \includegraphics[scale=0.45]{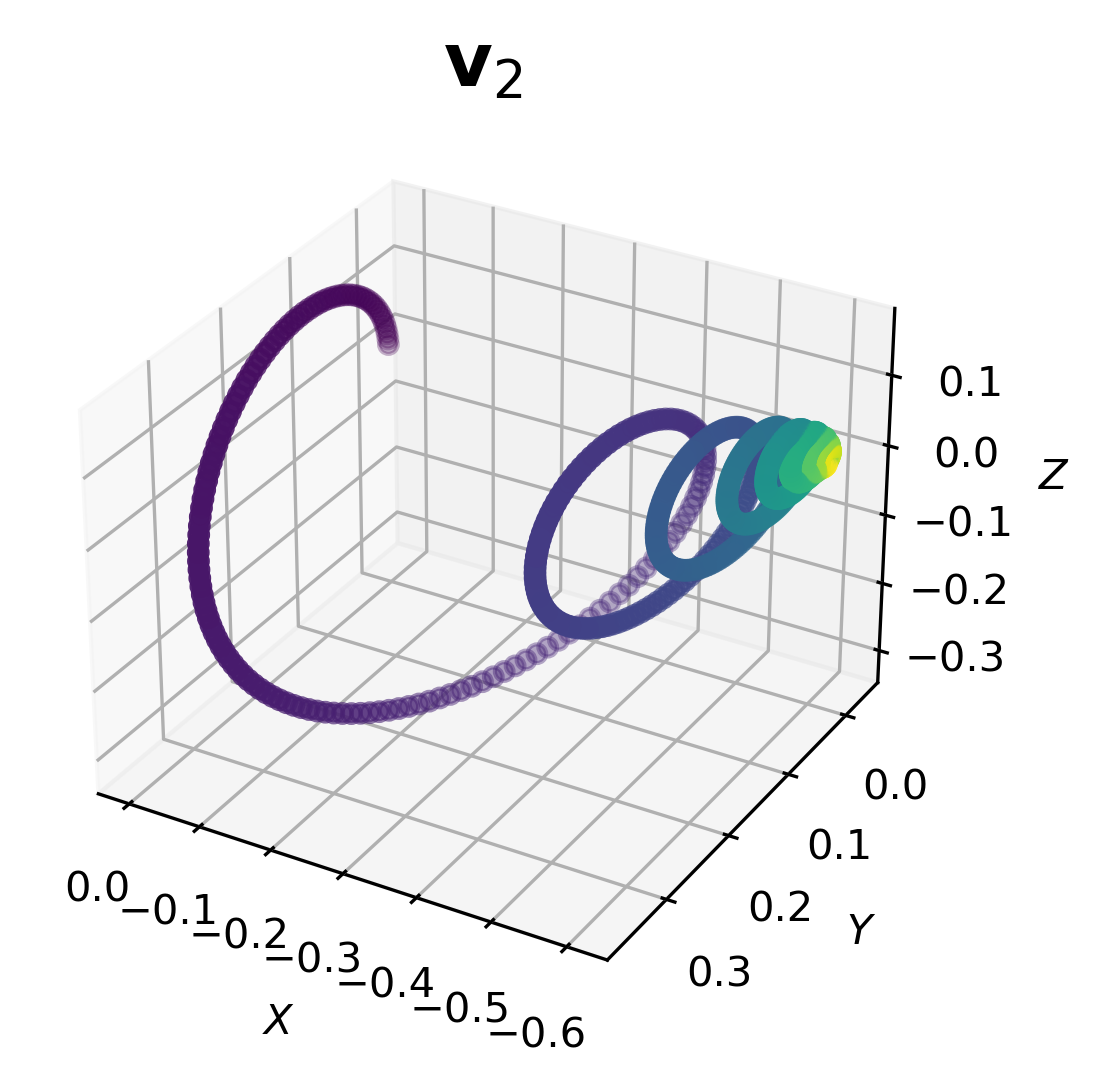}
  }
  \subfloat[]{
    \includegraphics[scale=0.45]{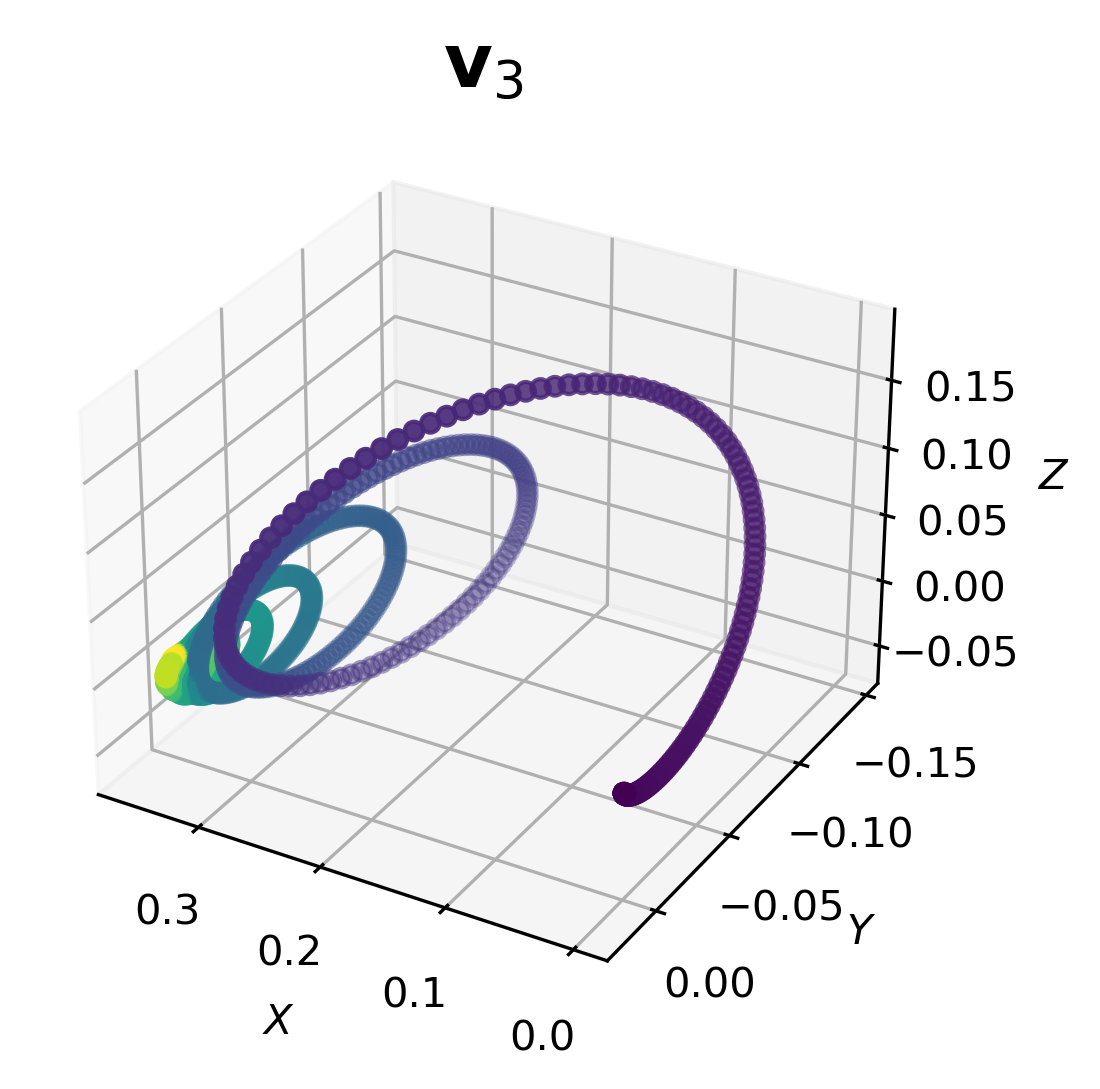}
  }
  \caption{\label{vector representation} The complete trajectories of
    the vector representations of Eq. (\ref{vgamma}) for our four Kraus
    operators.  Here we use
    $\rho_{S}\left(0\right)=\left|1\right\rangle \left\langle 1\right|,$
    $\omega_{c}=8\pi/\tau,$ $T=\hbar\omega_{c}/k_{B},$ $\eta=0.05,$ and
    $\Omega\left(t\right)$ as given by Eq. (\ref{cdd}), with $n=2.$ Time
    evolution proceeds from purple (darker) to yellow (lighter).}

\end{figure}

In view of the correspondence given by Eqs. (\ref{angles0}-\ref{angles3}),
a convenient way to represent the Kraus operators of Eqs. (\ref{K0}-\ref{K3})
is, therefore, by four respective Bloch vectors in three dimensions,
which we choose as the following:
\begin{eqnarray}
\mathbf{v}_{\gamma}\left(t\right) & \equiv & \sqrt{D_{\gamma}\left(t\right)}\left[\mathbf{\hat{x}}\sin\theta_{\gamma}\cos\varphi_{\gamma}\right.\nonumber \\
 &  & \left.+\mathbf{\hat{y}}\sin\theta_{\gamma}\sin\varphi_{\gamma}+\mathbf{\hat{z}}\cos\theta_{\gamma}\right],\label{vgamma}
\end{eqnarray}
for $\gamma=0,1,2,3.$ Because of Eqs. (\ref{normalization}) and
(\ref{normalization-tilde}), it follows that
\begin{eqnarray*}
\left|\mathbf{v}_{\gamma}\left(t\right)\right| & = & \sqrt{D_{\gamma}\left(t\right)},
\end{eqnarray*}
for $\gamma=0,1,2,3.$ Furthermore, we can derive from Eqs. (\ref{rhoIS(t)-1})
and (\ref{diag}) that
\begin{eqnarray*}
\sum_{\gamma=0}^{3}D_{\gamma}\left(t\right) & = & 1.
\end{eqnarray*}
As illustrations of the Kraus operators as three-dimensional vectors,
Fig. \ref{vector representation} shows the complete trajectories
of the vector representations of Eq. (\ref{vgamma}) for our four
Kraus operators in the case in which we use $\rho_{S}\left(0\right)=\left|1\right\rangle \left\langle 1\right|,$
$\omega_{c}=8\pi/\tau,$ $T=\hbar\omega_{c}/k_{B},$ $\eta=0.05,$
and $\Omega\left(t\right)$ as given by Eq. (\ref{cdd}), with $n=2.$

The representation prescribed here shows the whole dynamics of the
$\sqrt{\mathrm{SWAP}}$ gate under the effects of dephasing mitigated
by the procedure of dynamical decoupling. We advance this visualization
recipe in the present context because the context, per se, is relevant.
However, we also point out that we could apply the principles involved
here for other qubit dynamics, although a graphical representation
might usually be unpractical for qudit systems, as their state representation
in terms of generalized Bloch vectors already indicates~\cite{Bertlmann_2008}.

\section{Quantum Process Tomography in the language of the Quantum Operational-Probabilistic Theory}
\label{sec:opts-quantum-process}

Here we make the connection of the formalism just presented with the quantum OPT framework, which we briefly review in Appendix \ref{sec:opts}. To reconstruct the
Kraus operators (in other words, the dynamics) from measurement frequencies
alone, we can measure the process matrix defined in Eq. (\ref{Mnumu}) by means of QPT~\cite{qpt}. We wish to illustrate this technique using a simple OPT, with Positive Operator-Valued Measure (POVM) measurements~\cite{michaelnielsen2011}. This subject has been heavily discussed in
the literature, and we point to Refs. \cite{Mohseni_Rezakhani_Lidar_2008,
  Gutoski_Johnston_2014,
  Torlai_Wood_Acharya_Carleo_Carrasquilla_Aolita_2020} for a broad
analysis of QPT, as well as
Refs. \cite{Altepeter_James_Kwiat, Henao_Restrepo} for Quantum State
Tomography (QST).

We describe the evolution of given preparations
\(\rho_{i}(0)\) by a generic noisy gate \(\mathcal{M}\) that acts on a
qubit, of which we suppose having no knowledge. The index on the preparation tells
us which initial condition we choose among Bloch vectors in the
following set:
\(\{\boldsymbol{0}, \hat{\mathbf{x}}, \hat{\mathbf{y}},
\hat{\mathbf{z}}\}\).

As in usual QST, we need a set of measurement operators to apply on the qubit. We choose the usual set
\begin{eqnarray}
  \label{eq:Es}
  E_0 &=& \op{0_x}{0_x}, E_1 = \op{1_x}{1_x},\\
  E_2 &=& \op{0_y}{0_y}, E_3 = \op{1_y}{1_y},\\
  E_4 &=& \op{0_z}{0_z},\text{ and } E_5 = \op{1_z}{1_z}.
\end{eqnarray}
A given element of this set is \(E_k\). Notice that
\(E_{2j} + E_{2j+1}\) are all equal to the identity, for \(j = 0,1,2\).

We describe the quantum process using its Kraus representation,
\begin{eqnarray}
  \mathcal{M}(\cdot) = \sum_{\alpha=0}^{3} K_{\alpha}(\cdot)K_{\alpha}^\dagger\label{eq:process_as_kraus},
\end{eqnarray}
with the condition
\begin{eqnarray}
  \sum_{\alpha=0}^{3} K_{\alpha}^\dagger K_{\alpha}=\Id\label{eq:kraus_cond}.
\end{eqnarray}

Each probability is associated with a circuit as shown in
Fig. \ref{fig:closedcirc} of Appendix \ref{sec:opts}, with the detail that, as we do not control
our process, \(\mathcal{M}_j = \mathcal{M}, \forall j\). This
probability can be written as
\begin{eqnarray}
  p_{i, k} &=& \tr[\mathcal{M}(\rho_i(0))E_{k}]\nonumber\\
           &=& \sum_{\alpha=0}^{3}
               \tr[K_{\alpha}\rho_i(0)K_{\alpha}^\dagger E_k]\label{eq:p_ik}.
\end{eqnarray}
We represent each Kraus operator as a linear combination of the same
\(\sigma_{\mu}\) used in Eq. \eqref{UI(t)}:
\begin{eqnarray}
  K_{\alpha} = \sum_{\mu=0}^{3} k^{\alpha}_{\mu}\sigma_{\mu}. \label{eq:stokes_kraus}
\end{eqnarray}
Thus, we can write
\begin{eqnarray}
  p_{i, k} &=& \sum_{\alpha=0}^{3}\sum_{\mu=0}^{3}\sum_{\nu=0}^{3}
               \tr[k^{\alpha}_{\mu}k^{\alpha\ast}_{\nu}\sigma_{\mu}\rho_i(0)\sigma_{\nu}
               E_k]
               \nonumber\\
           &=& \sum_{\mu=0}^{3}\sum_{\nu=0}^{3}
               M_{\mu\nu}\tr[\sigma_{\mu}\rho_i(0)\sigma_{\nu}E_k] \label{eq:prob_M_trace}
\end{eqnarray}
where, by inspection, we notice that
\begin{eqnarray}
  M_{\mu\nu} = \sum_{\alpha=0}^{3} k^{\alpha}_{\mu}k^{\alpha\ast}_{\nu} \label{eq:M_stokes}
\end{eqnarray}
is the same matrix defined in Eq. (\ref{Mnumu}). An auxiliary
definition is
\begin{eqnarray}
  Q^{ik}_{\mu\nu} := \tr[\sigma_{\mu}\rho_i(0)\sigma_{\nu}E_k] \label{eq:Qikmunu},
\end{eqnarray}
which does not depend on the measurement frequencies themselves, but on the choice
of measurement operators. The index \(k\) runs from zero to five, as we have six
measurement operators, and the index \(i\) runs from zero to three, representing
the four possible initial conditions of the Bloch vector. With these
definitions, we obtain
\begin{eqnarray}
  p_{i,k} &=& \sum_{\mu=0}^{3}\sum_{\nu=0}^{3}M_{\mu\nu}
              Q^{ik}_{\mu\nu}\nonumber\\
          &=& \tr[M^TQ^{ik}].\label{eq:pMQ}
\end{eqnarray}
If we define \(\ket{M}\) as a column vector by stacking columns of
\(M\), and \(\ket{Q^{ikT}}\) by stacking columns of \((Q^{ik})^T\), then 
\begin{eqnarray*}
  p_{i,k} = \ip{Q^{ikT}}{M}.
\end{eqnarray*}
Notice that we can define a 24-row vector \(\ket{p}\) as
\begin{eqnarray}
  \ket{p} = \sum_{k=0}^{5}\sum_{i=0}^{3} p_{i,k}\ket{4k+i},
  \label{eq:p_as_ket}
\end{eqnarray}
where \(4k+i\) indexes the vector row.  In the same fashion, we define a
\(24\times16\) superoperator \(\mathcal{A}\) as
\begin{eqnarray}
  \label{eq:A_superop}
  \mathcal{A} \equiv \sum_{k=0}^{5}\sum_{i=0}^{3} \op{4k+i}{Q^{ikT}}.
\end{eqnarray}
These definitions lead us to
\begin{align}
  \label{eq:superop_linear_sys}
  \mathcal{A}\ket{M} = \ket{p},
\end{align}
from which it follows that
\begin{align}
  \label{eq:M_tomo}
  \ket{M} = (\mathcal{A}^T\mathcal{A})^{-1}\mathcal{A}^T\ket{p}.
\end{align}

\begin{figure}
  \centering
  \subfloat[]{
    \includegraphics[scale=0.35]{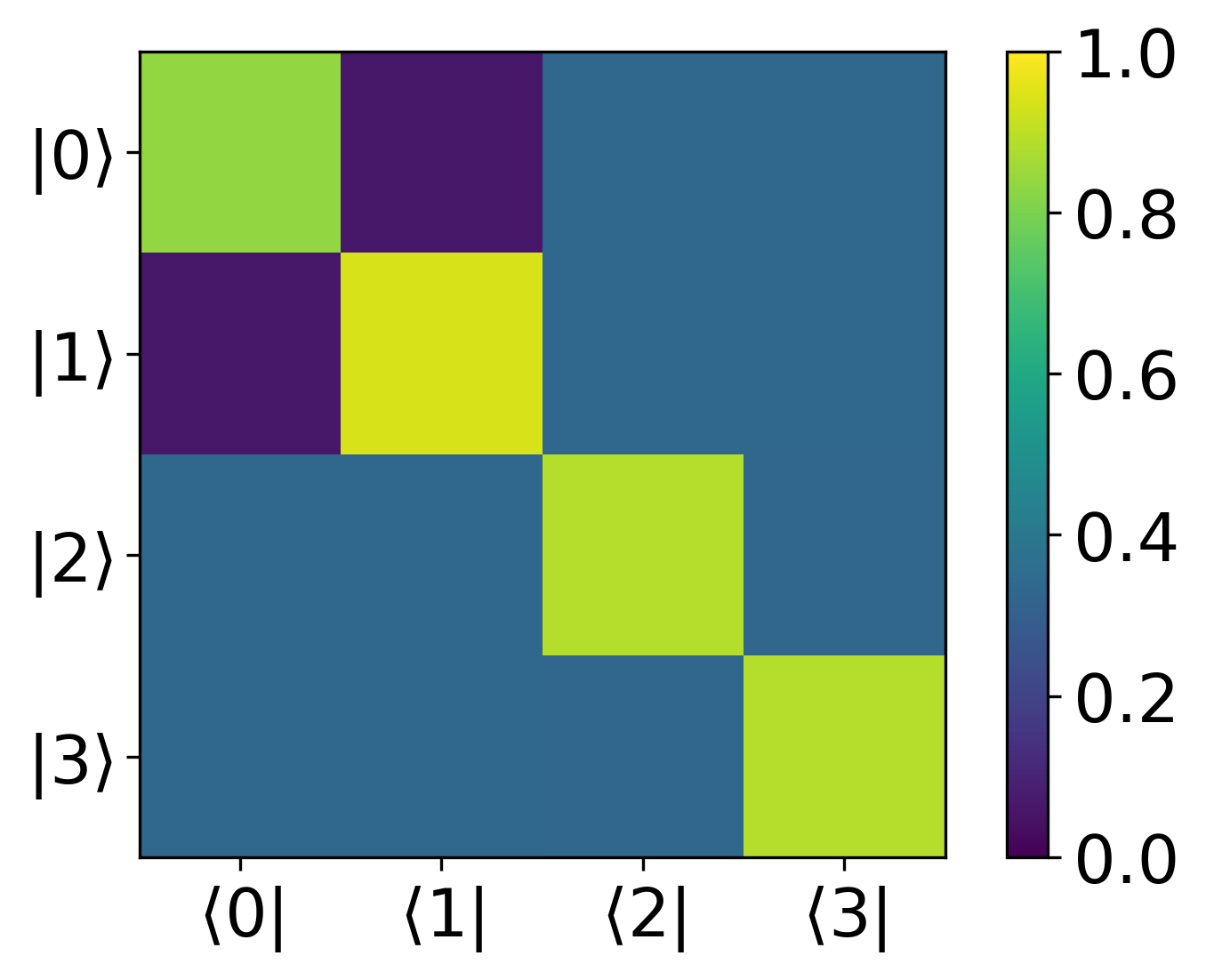}
    \label{fig:m_real}
  }
  \subfloat[]{
    \includegraphics[scale=0.35]{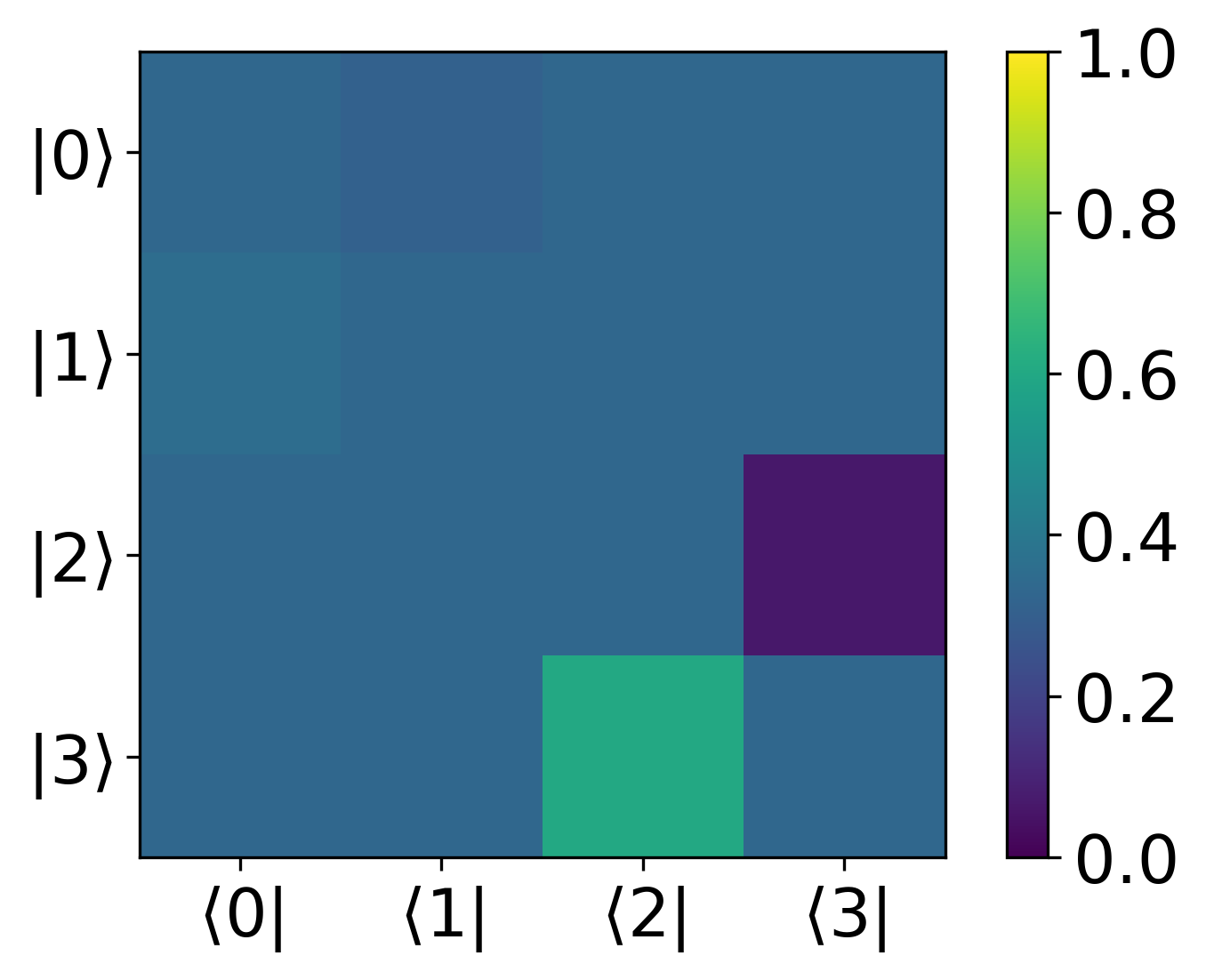}
    \label{fig:m_imag}
    }
  \\
  \subfloat[]{
    \includegraphics[scale=0.35]{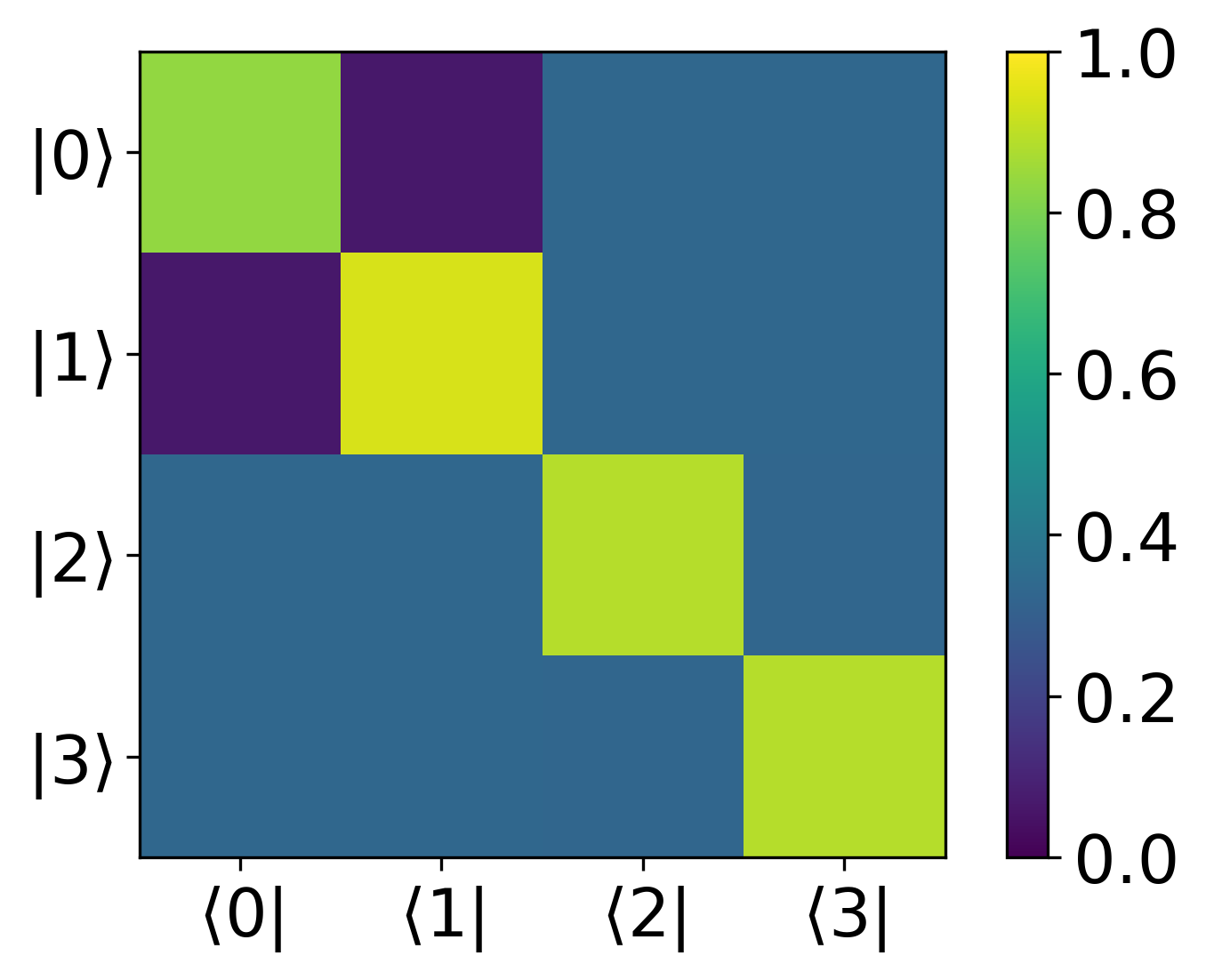}
    \label{fig:m_tomo_real}
  }
  \subfloat[]{
    \includegraphics[scale=0.35]{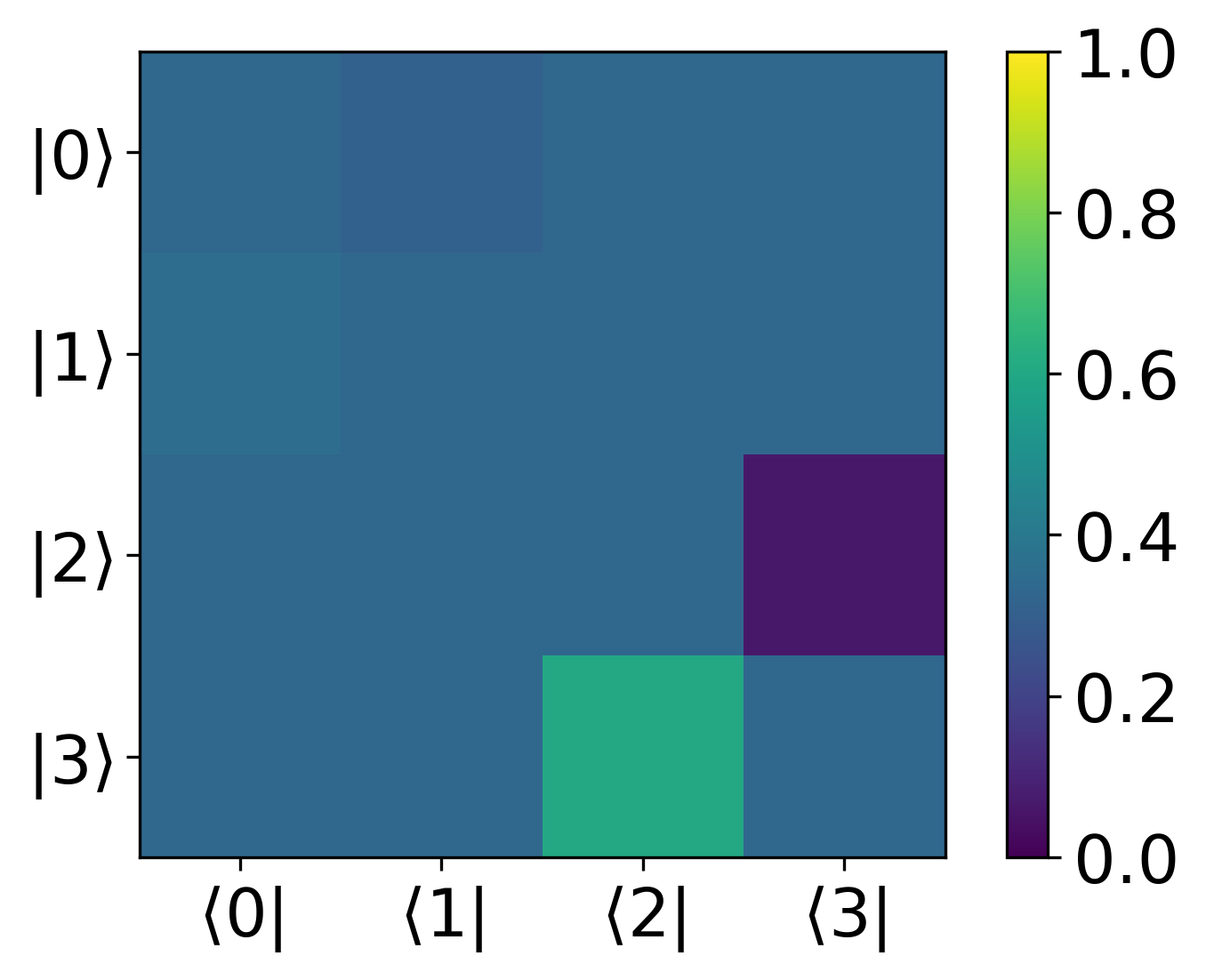}
    \label{fig:m_tomo_imag}
  }
  \caption{Figures (a) and (b):\ignore{Hinton} diagrams \(\mathrm{Re}[M]\) and
    \(\mathrm{Im}[M]\), obtained by solving
    Eq. \eqref{trajectory}. Figures (c) and (d):\ignore{Hinton} diagrams of the real
    and imaginary parts of tomographed \(M\). In all figures,
    \(t=\tau\), \(\omega_c = 8\pi/\tau\), \(T=\hbar\omega_c/k_B\),
    \(\eta=0.05\) and \(\Omega(t)\) as given by Eq. \eqref{cdd} with
    \(n=2\).}
  \label{fig:m_vs_tomo}
\end{figure}

As a self-consistency check, a comparison between \(M\) obtained by solving Eq. (\ref{Master}) or
(\ref{trajectory}) numerically, then solving Eq. (\ref{M-eqs}), and the one obtained by using the measurement frequencies given by Eq. (\ref{eq:M_tomo}) can be seen
in Fig. \ref{fig:m_vs_tomo}. Figures \subref*{fig:m_real} and
\subref*{fig:m_imag} are \ignore{Hinton} diagrams of \(\mathrm{Re}[M]\) and
\(\mathrm{Im}[M]\), respectively, obtained by Eq. (\ref{M-eqs}). This \(M\) gives us probabilities via
Eq. \eqref{eq:p_ik}. These probabilities have been used in the
method discussed above to reobtain \(M\) through Eq. (\ref{eq:M_tomo}). Figures \subref*{fig:m_tomo_real} and
\subref*{fig:m_tomo_imag} are diagrams of real and imaginary parts of this 
tomographed \(M\), Eq. (\ref{eq:M_tomo}). In the calculations presented in Fig. \ref{fig:m_vs_tomo} we used
\(t=\tau\), \(\omega_c = 8\pi/\tau\), \(T=\hbar\omega_c/k_B\),
\(\eta=0.05\) and \(\Omega(t)\) as given by Eq. \eqref{cdd} with
\(n=2\).

As the matrices obtained by the procedures associated with Eqs. (\ref{M-eqs}) and (\ref{eq:M_tomo}) are the same, we can de facto visualize the dynamics of
a system by knowing only the frequencies of each measurement, given each
initial state, since we can convert the process matrix \(M\)  obtained by QPT, via Eq. (\ref{eq:M_tomo}), to a Kraus
representation as explained in Sec. \ref{sec:visu-kraus-oper}.

\section{Conclusion}
\label{sec:conclusion}

In summary, we have considered the entangling universal quantum gate
$\sqrt{\mathrm{SWAP}}$ under the perturbation of a dephasing bath
using a spin-boson Hamiltonian. Under the guidance of quantum operational-probabilistic theories, we argue for the importance of the calculation
of Kraus operators, required to implement the completely-positive
maps that are essential in the context of these theories. Usually
one expects us to provide some means to overcome most of the deleterious
effects of the decoherence promoted by the environmental noise. We
accordingly have chosen continuous dynamical decoupling as such a protective
control procedure and have calculated a set of Kraus operators for the
residual dephasing. Inspired by the Bloch-vector trajectories which
describe the reduced-state evolution of the system, in this case starting from
an initial condition, we have prescribed a three-dimensional vector
representation for the whole residual-noise dynamics, proving that
such curves are sufficient to represent their corresponding Kraus-operator
histories. We once again emphasize the relevance of the fact that
the four Kraus-vector trajectories represent the whole dynamics of
the qubit system, independently of its initial state. Finally, we have discussed the 
procedure to obtain the process matrix and, thus, the Kraus operators and their visual representation, in the 
OPT framework, where only a set of POVM measurement frequencies are assumed known.

\appendix

\section{Operational-Probabilistic Theories}
\label{sec:opts}
References~\cite{PhysRevA.81.062348,PhysRevA.84.012311} present the main
ideas behind an OPT, with
changes proposed by Ref.~\cite{Oreshkov2015} to turn such a theory
symmetric under time reversal. Reference~\cite{morazotti_reversao_2018},
specially Chapter 4, explains how we can apply such a theory to a
Stern-Gerlach weak-value problem.

An OPT is a theory that describes possible experiments to be done with
physical apparatuses and gives predictions for outcomes of said experiments.
It employs two basic notions: \emph{systems} and \emph{operations}. An
operation represents a use of a physical device that connects systems
to systems. Examples include a set of mirrors in an optical
experiment or magnets in a Stern-Gerlach apparatus. Systems, as now may be clear,
are the particles subject to the experimentation, as the atoms leaving the
oven in a Stern-Gerlach setup.

Mathematically, systems are described by the usual Hilbert spaces.
Operations are maps from systems to systems. As they are physical maps,
they must be completely positive (CP) and trace preserving (TP). If a map is both
CP and TP it is called a completely-positive and trace-preserving (CPTP) map.

\begin{figure}
  \centering
  \subfloat[]{
    $$\Qcircuit @C=0.5em @R=0em{\prepareC{\{\rho_i\}} & \qw & A }$$
    \label{fig:preparations}
  }
  \quad
  \subfloat[]{
    $$\Qcircuit @C=.5em @R=0em {\lstick A & \gate{\{\mathcal{M}_j\}} & \qw & B}$$
    \label{fig:operations}
  }
  \quad
  \subfloat[]{
   $$\Qcircuit @C=0.5em @R=0em{\lstick B & \measureD{\{E_k\}}}$$
    \label{fig:measurements}
  }
  \caption{\protect\subref{fig:preparations} is a preparation $\rho$, with
    possible outcomes indexed by $i$; \protect\subref{fig:operations} is an
    operation $\mathcal{M}$, with possible outcomes indexed by $j$;
    and \protect\subref{fig:measurements} is a measurement $E$, with possible
    outcomes indexed by $k$.}
   \label{fig:opt}
\end{figure}
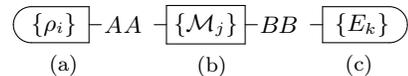

An OPT has the advantage that it may be understood through diagrams as
shown in Fig. \ref{fig:opt}. In Fig.  \subref*{fig:operations} the box
labeled by \({\cal M}\) is an operation, where the index \(i\)
represents an element of the set of outcomes. In this case, the wires
\(A\) and \(B\) are systems. Systems evolve from left to right (e.g., we
operate over the input system \(A\) with an operation \({\cal M}\) and
arrive at the output system \(B\)).

We may not care about our system before a given operation (in which case
we hereon call such system ``identity''). In terms of cooking, we
usually do not care about, say, potatoes proceeding before cooking; we
just wash and cook them (see Ref.~\cite{coecke_quantum_2010} for the
culinary example). This is called a \emph{preparation}: we prepare the
system so it is suitable to be studied in an experimental setting. In
the diagram representation, we draw it with a curved input, as shown in
Fig. \subref*{fig:preparations}. We may also not care about the system after
a given operation (e.g., after eating a meal); this is called
\emph{measurement} and we represent it with a curved output, as shown in
Fig. \subref*{fig:measurements}.

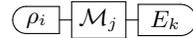
\begin{figure}
  \centering
  \[
    \Qcircuit @C=.5em @R=0em{ \prepareC {\rho_i} & \gate{{\cal M}_j} & \measureD{E_k}
    }
  \] 
  \caption{Circuits represent probabilities.}
  \label{fig:closedcirc}
\end{figure}

We can compose preparations, operations and measurements in a closed (no
open wires), non-cyclic (output from later operations cannot be used as
input for previous operations) fashion, called \emph{circuit} as done in
Fig. \ref{fig:closedcirc}. In an OPT, each circuit is associated with a
probability. In other words, if we have a set of preparations
\(\{\rho_i\}_{i\in{\cal O}}\), operations
\(\{{\cal M}_j\}_{j\in{\cal Q}}\), and measurements
\(\{E_k\}_{k\in{\cal R}}\), then
\begin{eqnarray}
    p(i,j,k|\{\rho_i\}_{i\in{\cal O}},
  \{{\cal M}_j\}_{j\in{\cal Q}},
  \{E_k\}_{k\in\mathcal R})
  \ge 0;\label{eq:p1}
\end{eqnarray}
\begin{eqnarray}
    \sum_{i\in {\cal O}}\sum_{j\in{\cal Q}}\sum_{k\in\mathcal R}
    p(i,j,k|\{\rho_i\}_{i\in{\cal O}},
    \{{\cal M}_j\}_{j\in{\cal Q}},
    \{E_k\}_{k\in\mathcal R})
   =1\nonumber\\ \label{eq:psum}.
\end{eqnarray}
For example, take a qubit in state \(\ket{0}\) (which is our
preparation) and operate a Hadamard gate \(H\) on it. Then, measure the
probability of the qubit being at \(\ket{0}\) or \(\ket{1}\). Our
preparation is, then, \(\rho = \op{0}{0}\). Our operation is \({\cal M}(\cdot) =
H(\cdot)H\).
Our measurement can be either \(E_1 = \op{0}{0}\) or \(E_2 =
\op{1}{1}\).
Consider the usual probability rule. Then, the probability \(p_1\)
of measuring \(E_1\) is
\begin{eqnarray}
  p_1 &=& \Qcircuit @C=.5em @R=0em{ \prepareC {\rho} & \gate{{\cal M}} & \measureD{E_1}}\nonumber\\
      &=& \tr[{\cal M}(\rho) E_1] \nonumber\\
      &=& \tr[H\op{0}{0}H\op{0}{0}]\nonumber\\
      &=& \frac{1}{2}\tr[(\op{0}{0}+\op{1}{1})\op{0}{0}]\nonumber\\
      &=& \frac{1}{2}.
\end{eqnarray}

In the same fashion, the probability \(p_2\) of measuring \(E_2\) is
\begin{eqnarray}
  p_2 &=& \Qcircuit @C=.5em @R=0em{ \prepareC {\rho} & \gate{{\cal M}} & \measureD{E_2}}\nonumber\\
      &=& \tr[{\cal M}(\rho) E_2] \nonumber\\
      &=& \tr[H\op{0}{0}H\op{1}{1}]\nonumber\\
      &=& \frac{1}{2}\tr[(\op{0}{0}+\op{1}{1})\op{1}{1}]\nonumber\\
      &=& \frac{1}{2}\\
  p_1 + p_2 &=& 1.  
\end{eqnarray}

These probabilities define an OPT. We make use of equivalence classes of
operations: if we have two operations, \(\{{\cal M}_{j}\}_{j\in{\cal Q}}\) and
\(\{{\cal N}_{j}\}_{j\in{\cal Q}}\), both with the same probability
distributions for all preparations and measurements described in the
theory, then they are considered the same operation. The same is valid
both for preparations and measurements.

The probability associated with each circuit is a variation of the usual
Hilbert-Schmidt inner product:
\begin{eqnarray}
  & &p(i,j,k|\{\rho_i\}_{i\in {\cal O}},
  \{{\cal M}_j\}_{j\in {\cal Q}},
  \{E_k\}_{k\in \mathcal R})\nonumber\\
  =& &\frac{\tr[{\cal M}_j(\rho_i)E_k]}
    {\sum_{i,j,k}\tr[{\cal M}_j(\rho_i)E_k]}
\end{eqnarray}
where we associate the operations, usually maps, to matrices via
isomorphism. If by happenstance there is an incompatibility of systems
between operations (say, the output of \({\cal M}\) is not compatible
with the input of \(E\) for every possible outcome of \({\cal M}\) and
\(E\)), said probability is zero.

Such change on probability is needed because, as Ref.~\cite{Oreshkov2015} points out, we need to define the
constraints for measurements, as they do not necessarily follow POVM's
constraints. Generally speaking, a measurement is a set of operators
\(\{E^A_k\}_{k\in{\cal R}}\) on system \(A\), each event named \emph{effect}, that
follows
\begin{eqnarray}
  E^A_{k} \leq \sum_{k\in\mathcal{R}}E^A_{k},\\
  \sum_{k\in{\cal R}} \tr[E^A_k] = d^A = \tr[\mathbb{I}^A].
\end{eqnarray}
An attentive reader notices that \(\{E^B_k\}_{k\in{\cal R}}\) \emph{may be} a
POVM, but does not need to.

A preparation, set of operators \(\{\rho^A_i\}_{i\in{\cal O}}\) in which each
operator is called \emph{state}, follows the density-matrix
constraint:
\begin{eqnarray}
  \rho^A_{i} \leq \sum_{i\in\mathcal{O}} \rho^A_{i},\\
  \sum_{i\in{\cal O}} \tr[\rho^A_i] = 1.
\end{eqnarray}

Operations, set of CP maps \(\{{\cal M}^{A\to B}_{j}\}_{j\in{\cal Q}}\) such that
each map is called a \emph{transformation}, have the constraint
\begin{eqnarray}
 \mathcal{M}^{A\to B}_j \leq \sum_{j\in\mathcal{Q}}\mathcal{M}^{A\to B}_j,\\
  \sum_{j\in{\cal Q}} \tr \left[ {\cal M}^{A\to B}_j
  \left( \frac{\mathbb{I}^A}{d^A} \right) \right] = 1.
\end{eqnarray}

\section{\label{=00005Cxi(t)}Derivation of \(\xi\left(t\right)\)}
Here we briefly outline how to get the result of Eqs. (\ref{exact})
and  (\ref{decaying coherence}). Using the evolution operator $U_{I}\left(t\right)$
in the interaction picture from Ref.~\cite{quiroga} adapted to a
single qubit, we can easily see that
\begin{eqnarray}
\mathrm{Tr}_{B}\left[U_{I}\left(t\right)\left|1\right\rangle \rho_{B}\left(0\right)\left\langle 1\right|U_{I}^{\dagger}\left(t\right)\right] & = & \left|1\right\rangle \left\langle 1\right|\label{00}
\end{eqnarray}
and
\begin{eqnarray}
\mathrm{Tr}_{B}\left[U_{I}\left(t\right)\left|2\right\rangle \rho_{B}\left(0\right)\left\langle 2\right|U_{I}^{\dagger}\left(t\right)\right] & = & \left|2\right\rangle \left\langle 2\right|,\label{11}
\end{eqnarray}
where $\rho_{B}\left(0\right)$ is given by Eq. (\ref{thermal}).
In the present context, of course, we are assuming $\Omega\left(t\right)=0$
in Eqs. (\ref{HI}), (\ref{UI}), and (\ref{UIevol}). A somewhat
more involved calculation gives
\begin{eqnarray}
\mathrm{Tr}_{B}\left[U_{I}\left(t\right)\left|1\right\rangle \rho_{B}\left(0\right)\left\langle 2\right|U_{I}^{\dagger}\left(t\right)\right] & = & \left|1\right\rangle \left\langle 2\right|\exp\left[\mathcal{P}\left(t\right)\right]\nonumber \\
 &  & \times\exp\left[\mathcal{Q}\left(t\right)\right],\label{01}
\end{eqnarray}
where
\begin{eqnarray}
\mathcal{P}\left(t\right) & \equiv & 4\sum_{s}\left|g_{s}\right|^{2}\frac{\cos\left(\omega_{s}t\right)-1}{\omega_{s}^{2}},\label{I}
\end{eqnarray}
\begin{eqnarray}
\mathcal{Q}\left(t\right) & \equiv & 8\sum_{s}\left\langle n_{s}\right\rangle \left|g_{s}\right|^{2}\frac{\cos\left(\omega_{s}t\right)-1}{\omega_{s}^{2}},\label{J}
\end{eqnarray}
 and we have used the very useful representation~\cite{gardiner_quantum_2004}:
\begin{eqnarray}
\rho_{B}\left(0\right) & = & \prod_{s}\frac{1}{\left\langle n_{s}\right\rangle }\frac{1}{\pi}\nonumber \\
 &  & \times\int d^{2}\alpha_{s}\,\exp\left(-\frac{\left|\alpha_{s}\right|{{}^2}}{\left\langle n_{s}\right\rangle }\right)\left|\alpha_{s}\right\rangle \left\langle \alpha_{s}\right|,\label{repr}
\end{eqnarray}
with
\begin{eqnarray}
\left\langle n_{s}\right\rangle  & \equiv & \frac{1}{\exp\left(\beta\hbar\omega_{s}\right)-1}\label{nsbar}
\end{eqnarray}
and $\left|\alpha_{s}\right\rangle $ is the usual coherent
state for the $s\text{th}$ mode of the boson bath. We can now use
the spectral density given in the continuum description of Eqs. (\ref{continuum})
and (\ref{ohmic}) to write Eq. (\ref{I}) as
\begin{eqnarray}
\mathcal{P}\left(t\right) & = & 4\eta\int_{0}^{\infty}d\omega\,\exp\left(-\frac{\omega}{\omega_{c}}\right)\frac{\cos\left(\omega t\right)-1}{\omega}.\label{sumtoint}
\end{eqnarray}
But,
\begin{eqnarray}
\frac{\cos\left(\omega t\right)-1}{\omega} & = & -\int_{0}^{t}dt^{\prime}\,\sin\left(\omega t^{\prime}\right),\label{trick}
\end{eqnarray}
so that now Eq. (\ref{sumtoint}) gives
\begin{eqnarray}
\mathcal{P}\left(t\right) & = & -2\eta\ln\left(1+\omega_{c}^{2}t^{2}\right).\label{almost}
\end{eqnarray}

Equation (\ref{J}) after using the spectral density given in the
continuum description of Eqs. (\ref{continuum}) and (\ref{ohmic})
and some algebraic manipulation can be put in this more convenient
form:
\begin{eqnarray}
\mathcal{Q}\left(t\right) & = & -8\eta\int_{0}^{\infty}d\omega\,\frac{\exp\left(-\frac{\omega}{\omega_{c}}\right)}{\exp\left(\beta\hbar\omega\right)-1}\frac{1-\cos\left(\omega t\right)}{\omega},\nonumber \\
\label{moreconvenient}
\end{eqnarray}
It is an invigorating exercise to review some techniques of mathematical
physics~\cite{georgearfken2011} to reexpress Eq. (\ref{moreconvenient})
in the form
\begin{eqnarray}
\mathcal{Q}\left(t\right) & = & \ln\left[\frac{\left|\left(\frac{k_{B}T}{\hbar\omega_{c}}+i\frac{k_{B}T}{\hbar}t\right)!\right|}{\left(\frac{k_{B}T}{\hbar\omega_{c}}\right)!}\right]^{8\eta}.\label{hairy}
\end{eqnarray}
From the initial qubit density operator, $\rho_{S}\left(0\right)=\left|\psi_{0}\right\rangle \left\langle \psi_{0}\right|$,
Eqs. (\ref{UI}), (\ref{pure noise}), (\ref{00}), (\ref{11}), (\ref{01})
with its Hermitian conjugate, and substitution of the results of Eqs.
(\ref{almost}) and (\ref{hairy}), we finally obtain
\begin{eqnarray}
\mathrm{Tr}_{B}\left[\rho_{I}\left(t\right)\right] & = & \left|c_{1}\right|^{2}\left|1\right\rangle \left\langle 1\right|\nonumber \\
 &  & +c_{1}c_{2}^{\ast}\xi\left(t\right)\left|1\right\rangle \left\langle 2\right|\nonumber \\
 &  & +c_{1}^{\ast}c_{2}\xi\left(t\right)\left|2\right\rangle \left\langle 1\right|\nonumber \\
 &  & +\left|c_{2}\right|^{2}\left|2\right\rangle \left\langle 2\right|,\label{finally}
\end{eqnarray}
where we have already recognized the definition of Eq. (\ref{decaying coherence}).
Thus we see that Eq. (\ref{finally}) is equivalent to Eq. (\ref{exact}).
\begin{acknowledgments}
  R.d.J.N. acknowledges support from Funda\c{c}\~{a}o de Amparo \`{a}
  Pesquisa do Estado de S\~{a}o Paulo (FAPESP), project number
  2018/00796-3, and also from the National Institute of Science and
  Technology for Quantum Information (CNPq INCT-IQ 465469/2014-0) and
  the National Council for Scientific and Technological Development
  (CNPq).  N. A. da C. M. acknowledges financial support from
  Coordena\c{c}\~ao de Aperfei\c{c}oamento de Pessoal de N\'ivel
  Superior (CAPES), project number 88887.339588/2019-00.
\end{acknowledgments}

\bibliography{refs} 
\end{document}